\documentclass{PoS}

\def\BE{\begin{equation}}
\def\EE{\end{equation}}
\def\BEA{\begin{eqnarray}}
\def\EEA{\end{eqnarray}}
\def\EL{\nonumber\\}

\title{A derivative-based approach for the leading order hadronic contribution to $g_\mu-2$}

\ShortTitle{A derivative-based approach to $g_\mu-2$}

\author{\speaker{Eric B. Gregory}\\
        Bergische Universit\"{a}t Wuppertal, 42097 Wuppertal, Germany\\
        J\"{u}lich Supercomputing Centre, 52428 J\"{u}lich, Germany\\
        E-mail: \email{gregory@uni-wuppertal.de}}

\author{Craig McNeile\\
        Plymouth University, PL4 8AA Plymouth, United Kingdom\\
        E-mail: \email{craig.mcneile@plymouth.ac.uk}}

\abstract{We describe a lattice approach to calculating the
  leading-order hadronic contribution to the anomalous magnetic moment of the muon. 
We employ lattice momentum derivatives, in both the spatial and
temporal directions, to determine the hadronic vacuum polarization
scalar at low momenta and construct a smooth, intregrable function in
this momentum region. 
The method is tested on one
hex-smeared Wilson-quark lattice ensemble with physical pion masses}

\FullConference{The 33rd International Symposium on Lattice Field Theory\\
		14 -18 July 2015\\
		Kobe International Conference Center, Kobe, Japan*}

\begin{document}

\section{Introduction}

The calculation of the anomalous magnetic moment of 
the muon $a_\mu=\frac{(g_\mu-2)}{2}$ is an important 
challenge, because a precise theoretical calculation from the standard model of
particle physics, which
differs from the experimental value, would be an indication of physics beyond the 
Standard Model. Indeed there is a current tension between the
experimental estimate for $a_\mu$, and the value predicted by
the standard model.
The hadronic contribution to 
$a_{\mu}$ is the dominant source of uncertainty. There are new
experiments at FNAL~\cite{Grange:2015fou} and J-PARC which
plan to reduce the experimental error on $a_\mu$, thus motivating
reducing the errors on the theoretical calculation.

In this work we report on the determination of the hadronic vacuum
polarization (HVP) contribution to $a_\mu$,
using a derivative based method.
The lattice determination of $a^{\rm HVP, LO}_{\mu}$ was pioneered by
Blum~\cite{Blum:2002ii}. 
Izubuchi~\cite{Taku} has reviewed
recent developments in calculating $a^{\rm HVP,LO}_{\mu}$ using lattice QCD.

The strategy used in this, and most previous lattice QCD calculations, is as follows.
First vector current
correlators are used to calculate the hadronic vacuum polarization (HVP) tensor in momentum space:
\begin{equation}
\label{HVP_tensor}
\Pi_{\mu\nu}(\hat{q}) = \sum_x e^{iq(\Delta x+\frac{a\hat{\mu}}{2})}
\langle J_\mu^{\rm CVC}(x_0)J_\nu^{\rm loc}(x)\rangle.
\end{equation}
Here $J_\nu^{\rm loc}$ is the local vector current and $ J_\mu^{\rm CVC}$ is the lattice conserved vector current which satisfies the Ward identity for the modifies momentum  $\hat{q}_\mu=\frac{2}{a}\sin\left(\frac{aq_\mu}{2}\right)$. 
From this one determines a HVP scalar
\begin{equation}
\label{HVP_scalar}
\Pi(s)\equiv \Pi_{\mu\nu}(\hat{q}) / T_{\mu\nu}(\hat{q}),
\end{equation}
with the momentum tensor $T_{\mu\nu}(\hat{q})\equiv \left(\hat{q}_{\mu}\hat{q}_{\nu} - \hat{q}^2\delta_{\mu\nu} \right)$, and $s=q^2$

The lowest-order contribution to $a_\mu^{\rm had}$ is given by
\begin{equation}
\label{integral}
a_\mu^{\rm had,LO} = \frac{\alpha}{\pi}\int^\infty_0ds \,f(s)\Pi_p(s),
\end{equation}
using the kernel function
\begin{equation}
f(s) = \frac{m^2_\mu s Z(s)^3 \left(1 - s Z(s)\right)}
{1 + m_\mu^2 s Z(s)^2},\;\;\;\;
{\rm where}
\;\;\;\;
Z=-\frac{s - \sqrt{s^2 + 4m^2_\mu s}}{2m_\mu^2 s}.
\end{equation}

In general only values of $\Pi(s)$ are known at discrete lattice momenta,
so some procedure is needed to determine a smooth function
$\Pi(s)$. In the past some groups have 
relied upon fitting a function, such as a vector 
meson dominance model, 
to the lattice values of $\Pi(s)$. This model-dependence  introduces potentially significant 
systematic effects~\cite{Golterman:2014ksa}.
A further challenge is that one cannot directly 
access the zero-momentum value of $\Pi(s)$ through equation~\ref{HVP_scalar}. 
This makes it harder to constrain the low-momentum values which contribute the most to the integral 
in equation~\ref{integral}.

We propose a moments-based method that addresses each of these concerns. 
We determine spatial and temporal momentum derivatives of $
T_{\mu\nu}(\hat{q})$. To estimate the spatial derivatives requires
additional
correlators to be measured.
From these momentum derivatives we can 
calculate that corresponding derivatives of the HVP scalar $\Pi(s)$.
We use Taylor expansions to interpolate $\Pi(s)$ to non-lattice values of $s$.
Our method produces a model-independent smooth curve for $\Pi(s)$ and allows direct access to the zero-momentum value
of $\Pi(s)$. This produces a high-precision determination of $\Pi(s)$ in 
the crucial low-momentum region of the integrand of (\ref{integral}).
De Rafael~\cite{deRafael:2014gxa} has shown that $a_\mu^{\rm had,LO}$ can be reconstructed 
from up to three derivatives of $\Pi_p(s)$.

\section{Outline of the method}

We begin by determining the HVP vector and the its first $N$ derivatives with respect to 
momenta $q_{{\alpha}_i}$ for $i=1,..,N$:
\BEA
\label{Pi_tensor}
\Pi_{\mu\nu}(\hat{q}) &=& \sum_x e^{iq(\Delta x+\frac{a\hat{\mu}}{2})}
\langle J_\mu^{\rm CVC}(x_0)J_\nu^{\rm loc}(x)\rangle\\
\frac{\partial^n \Pi_{\mu\nu}(\hat{q})}{\partial q_{\alpha_1}\cdots\partial q_{\alpha_n}} &=& 
i^n\sum_x
\left[
\prod_\rho^n
(\Delta x_{\alpha_\rho}+\frac{\delta_{\mu\alpha_\rho}}{2}) 
\right]
e^{iQ(\Delta x+\frac{a\hat{\mu}}{2})}
\langle J_\mu^{\rm CVC}(x_0)J_\nu^{\rm loc}(x)\rangle.
\EEA

We generally determine $N=8$ derivatives of $\Pi_{\mu\nu}$ using both
spatial and temporal moments, which we will see gives three
derivatives of $\Pi(s)$. Other groups,
e.g.~\cite{Chakraborty:2014mwa}, have used temporal moments.
However apart from the proposal in~\cite{deDivitiis:2012vs},
no other groups, to our knowledge, have taken advantage of the spatial moments.

First we
transform derivatives of $\Pi_{\mu\nu}$ with respect to q, to
derivatives with respect to $\hat{q}$. 
This is straightforward with the chain rule.
To determine derivatives of $\Pi(s)$ we again apply the chain rule.
Linear expressions relate derivatives of  $\Pi(s)$ and ${\Pi}_{\mu\nu}(q)$:
\begin{equation}
\label{linear_expression}
\frac{\partial^n \Pi_{\mu\nu}}{\partial q_{\alpha_1}\cdots\partial q_{\alpha_n}}
(q) 
= \sum_{m=0}^n{A_{\mu\nu}^{\{\alpha\}}}^n_m(q) \frac{d^m \Pi(s)}{d s^m}.
\end{equation}
The superscript $\{\alpha\}$ is shorthand for the set of indices $\alpha_1\cdots\alpha_n$. We will occasionally suppress the  $\{\alpha\}$  for readability.
Recursion expressions relate the $A^n_m$ to 
$
{A_{\mu\nu}}^0_0(q) = T_{\mu\nu}(q).
$
The $m=0$ terms are derivatives of $T_{\mu\nu}(q)$:
\BEA
{A_{\mu\nu}^{\{\alpha\}}}^n_0(q) &=&
 \partial_{\alpha_n}\cdots\partial_{\alpha_1}{A_{\mu\nu}}^0_0(q) \EL
&=&\partial_{\alpha_n}\cdots\partial_{\alpha_1}T_{\mu\nu}(q).
\EEA
Note that $T_{\mu\nu}(q)$ has only three non-zero derivatives:
\begin{equation}
{A_{\mu\nu}^{\{\alpha\}}}^n_0(q) =
\left\{
\begin{array}{lll}
T_{\mu\nu}(q)&=q_\mu q_nu - q^2\delta_{\mu\nu}& {\mbox{for }n=0}\\
\frac{\partial T_{\mu\nu}}{\partial q_{\alpha_1}} &= \delta_{\mu\alpha_1}q_\nu + \delta_{\nu\alpha_1}q_\mu - 2\delta_{\mu\nu}q_{\alpha_1}& {\mbox{for }n=1}\\
\frac{\partial^2 T_{\mu\nu}}{\partial q_{\alpha_1}\partial q_{\alpha_2}}&= \delta_{\mu\alpha_1}\delta_{\nu\alpha_2} + \delta_{\mu\alpha_2}\delta_{\nu\alpha_1} + 
2\delta_{\mu\nu}\delta_{\alpha_1\alpha_2}
& {\mbox{for } n=2}\\
\frac{\partial^n T_{\mu\nu}}{\partial q_{\alpha_1}\cdots\partial q_{\alpha_n}}&=0 & {\mbox{for }n<2}
\end{array}
\right.
\end{equation}
One finds also that the when $m=n$
\begin{equation}
{A_{\mu\nu}^{\{\alpha\}}}^n_n(q) = 
\left\{
\begin{array}{ll}
2q_{\alpha_n}{A_{\mu\nu}^{\{\alpha\}}}^{n-1}_{n-1} & \mbox{for } n < 3\\
0 & \mbox{for } n\geq 3,
\end{array}
\right.
\end{equation}
and, in general
\begin{equation}
\label{general_Amn}
{A_{\mu\nu}^{\{\alpha\}}}^n_m =
2q_{\alpha_n}{A_{\mu\nu}^{\{\alpha\}}}^{n-1}_{m-1} 
+ \partial_{q_{\alpha_n}}{A_{\mu\nu}^{\{\alpha\}}}^{n-1}_{m}.
\end{equation}
The expressions for ${A_{\mu\nu}^{\{\alpha\}}}^n_m$ tend to have a large number of terms.  We have a script that 
generates algebraic and C code expressions for these.

For non-zero momentum we can now compute $\frac{\partial^n \Pi_{\mu\nu}(\hat{q})}{\partial q_{\alpha_1}\cdots\partial q_{\alpha_n}}$ by solving the linear system (\ref{linear_expression}).

\begin{figure}
\begin{center}
\includegraphics[scale=0.2]{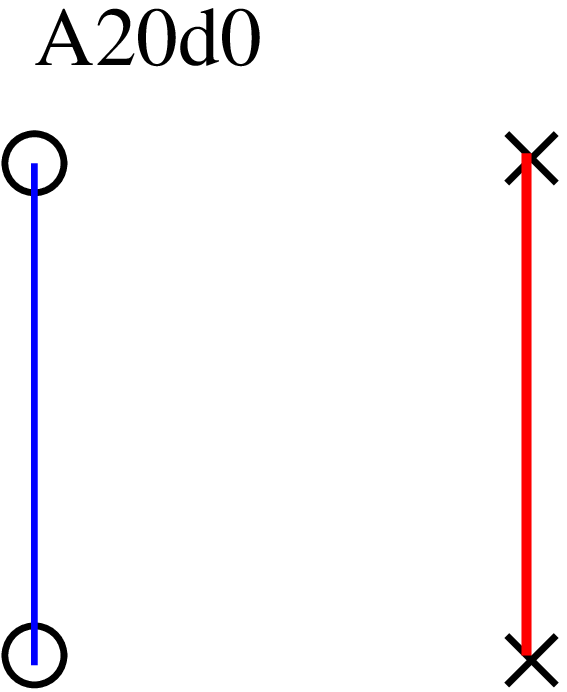}\hspace{0.5in}
\includegraphics[scale=0.2]{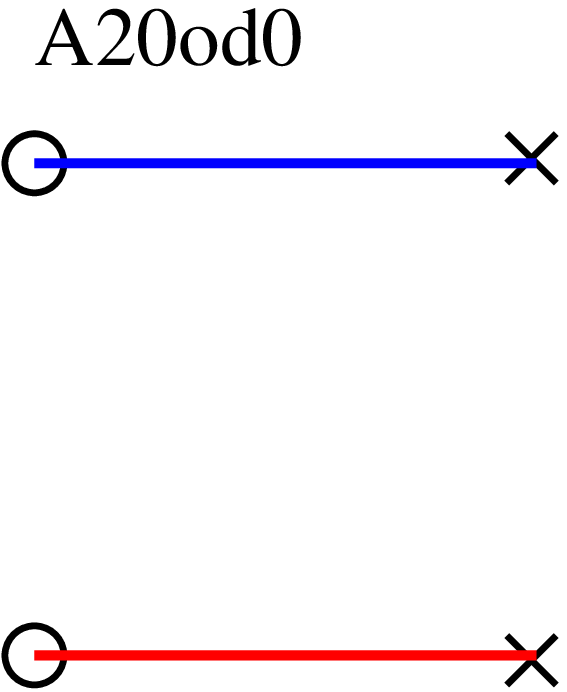}\vspace{-0.2in}
\end{center}
\caption{\label{fig:A20comb} Graphical depiction of contributing $A^2_0$ index combinations. Circles represent the $\mu$ and $\nu$ indices, crosses represent $\alpha$ indices. Colored bars indicate the connected indices have the same value.}
\end{figure}

\begin{figure}
\begin{center}
\includegraphics[scale=0.185]{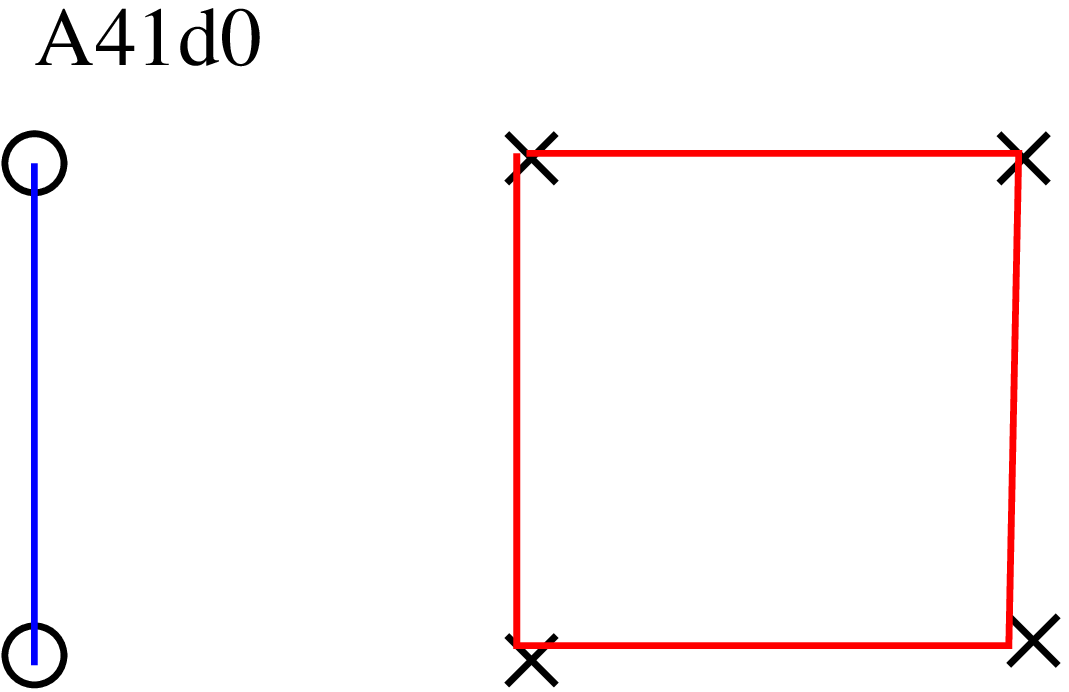}\hspace{0.4in}
\includegraphics[scale=0.185]{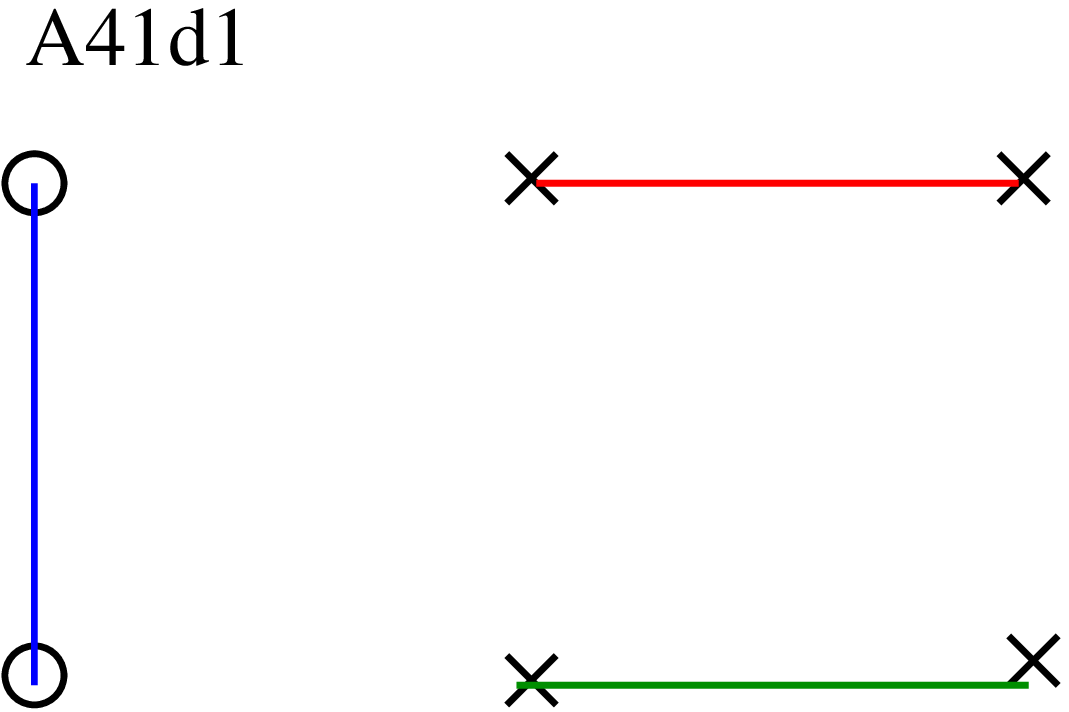}\hspace{0.4in}
\includegraphics[scale=0.185]{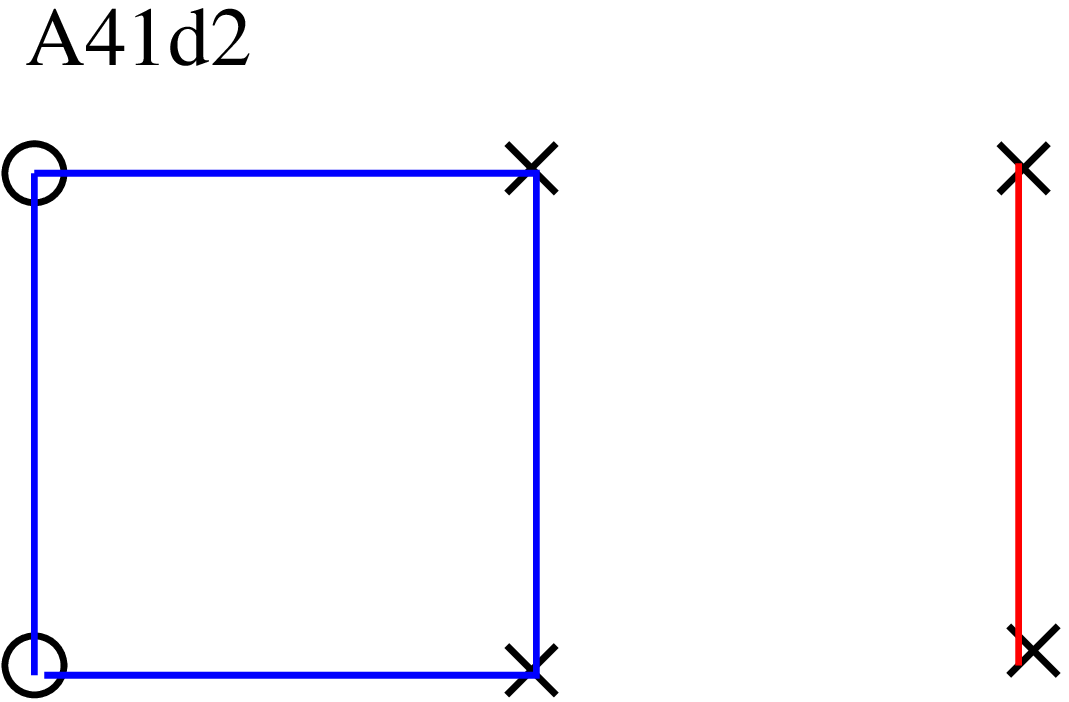}\hspace{0.4in}
\includegraphics[scale=0.185]{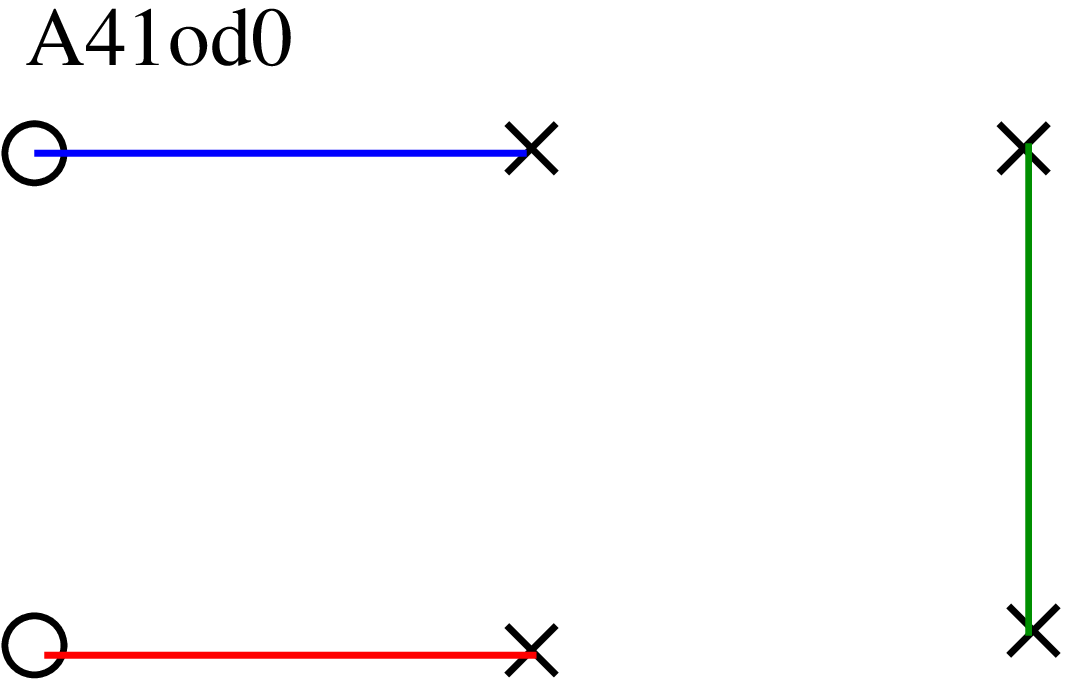}\hspace{0.4in}
\includegraphics[scale=0.185]{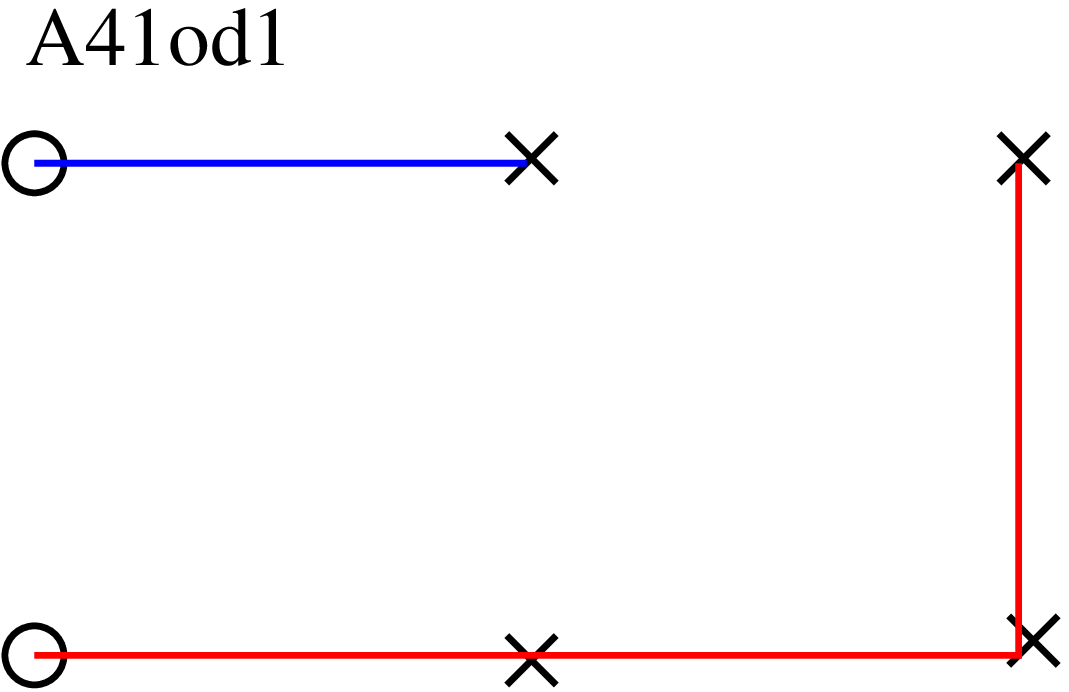}\vspace{-0.2in}
\end{center}
\caption{\label{fig:A41comb} Graphical depiction of contributing $A^4_1$ index combinations.}
\end{figure}

\begin{figure}
\begin{center}
\includegraphics[scale=0.135]{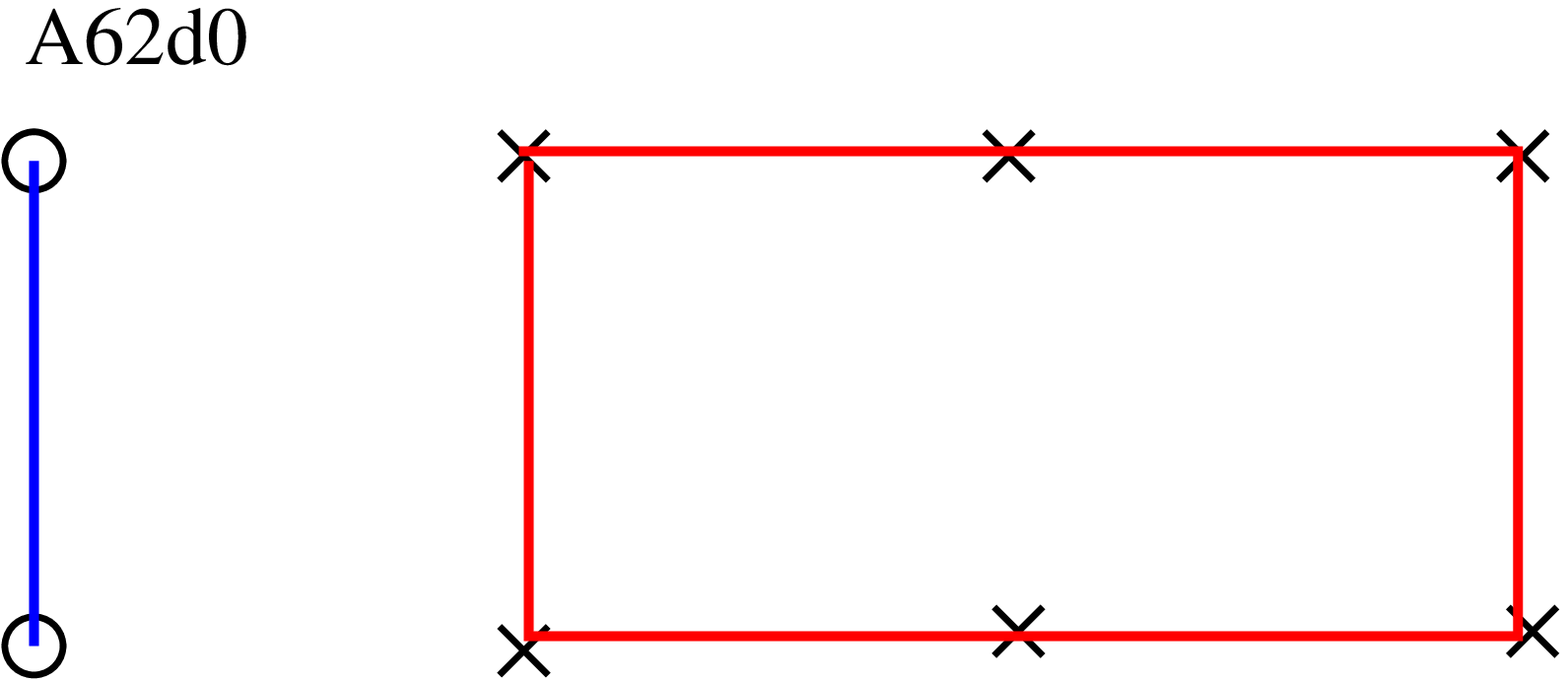}\hspace{0.35in}
\includegraphics[scale=0.135]{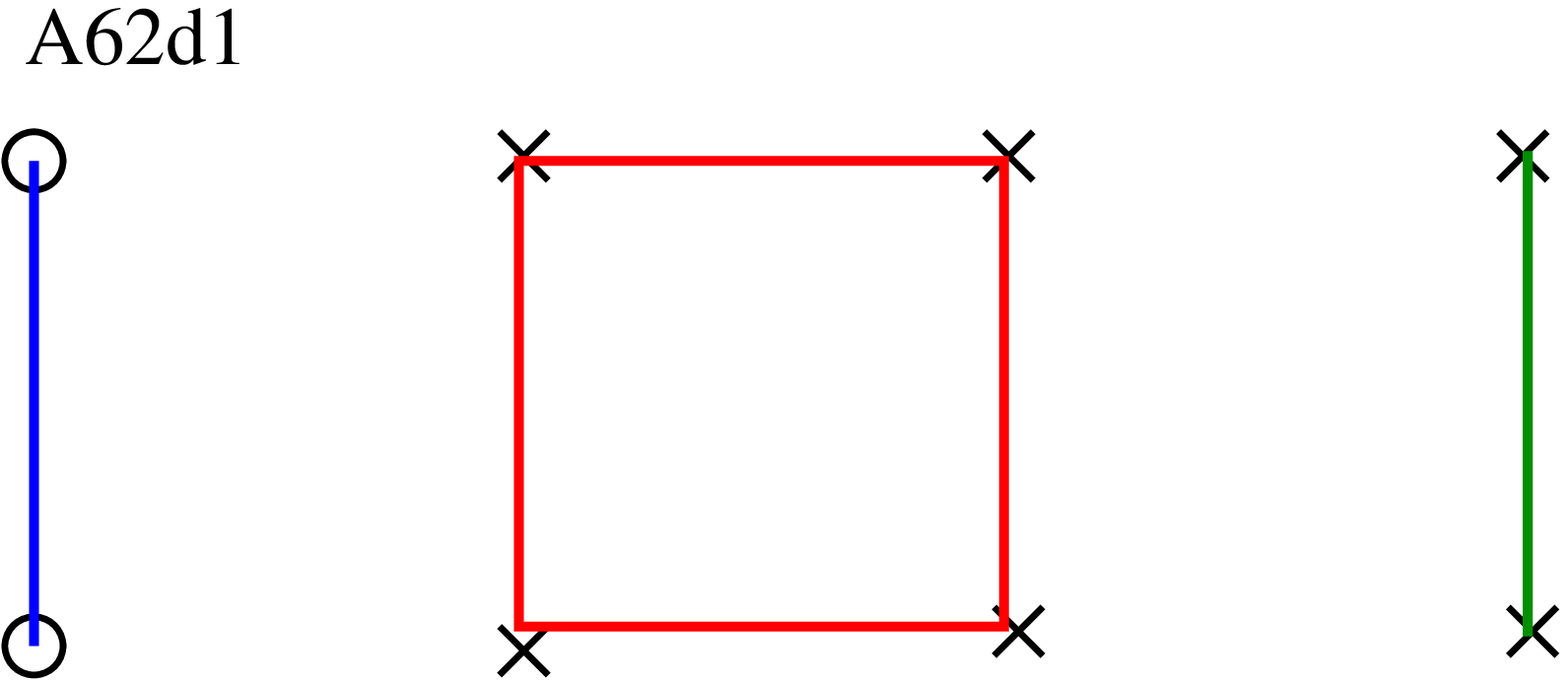}\hspace{0.35in}
\includegraphics[scale=0.135]{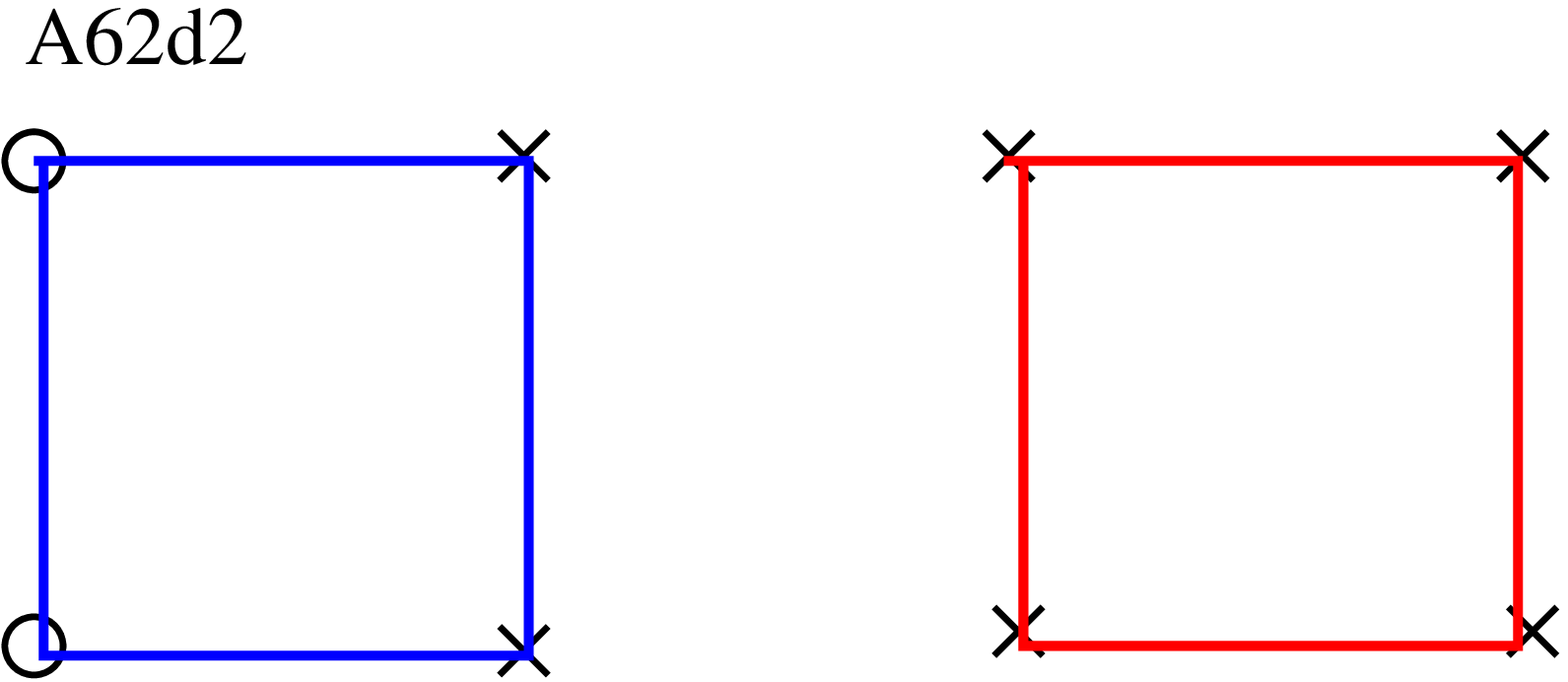}\hspace{0.35in}
\includegraphics[scale=0.135]{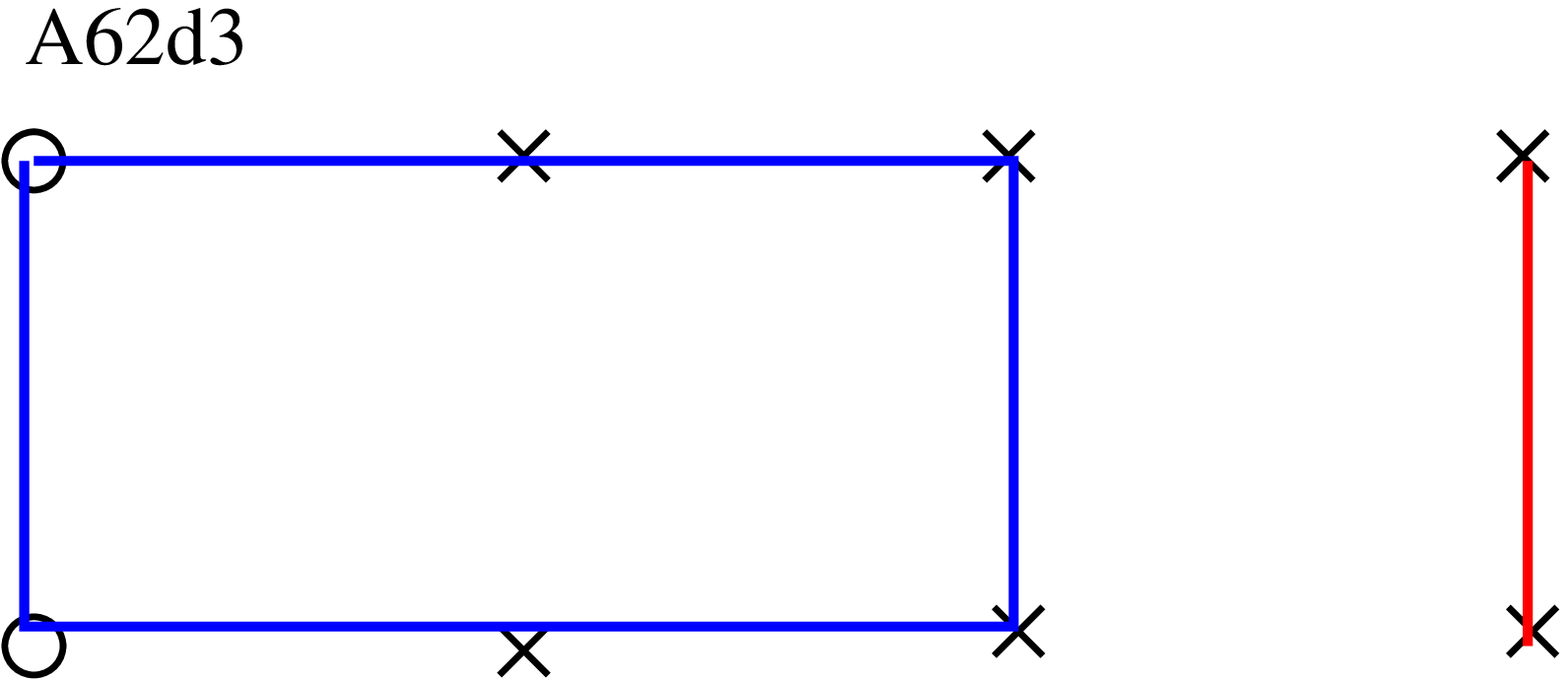}\hspace{0.35in}
\includegraphics[scale=0.135]{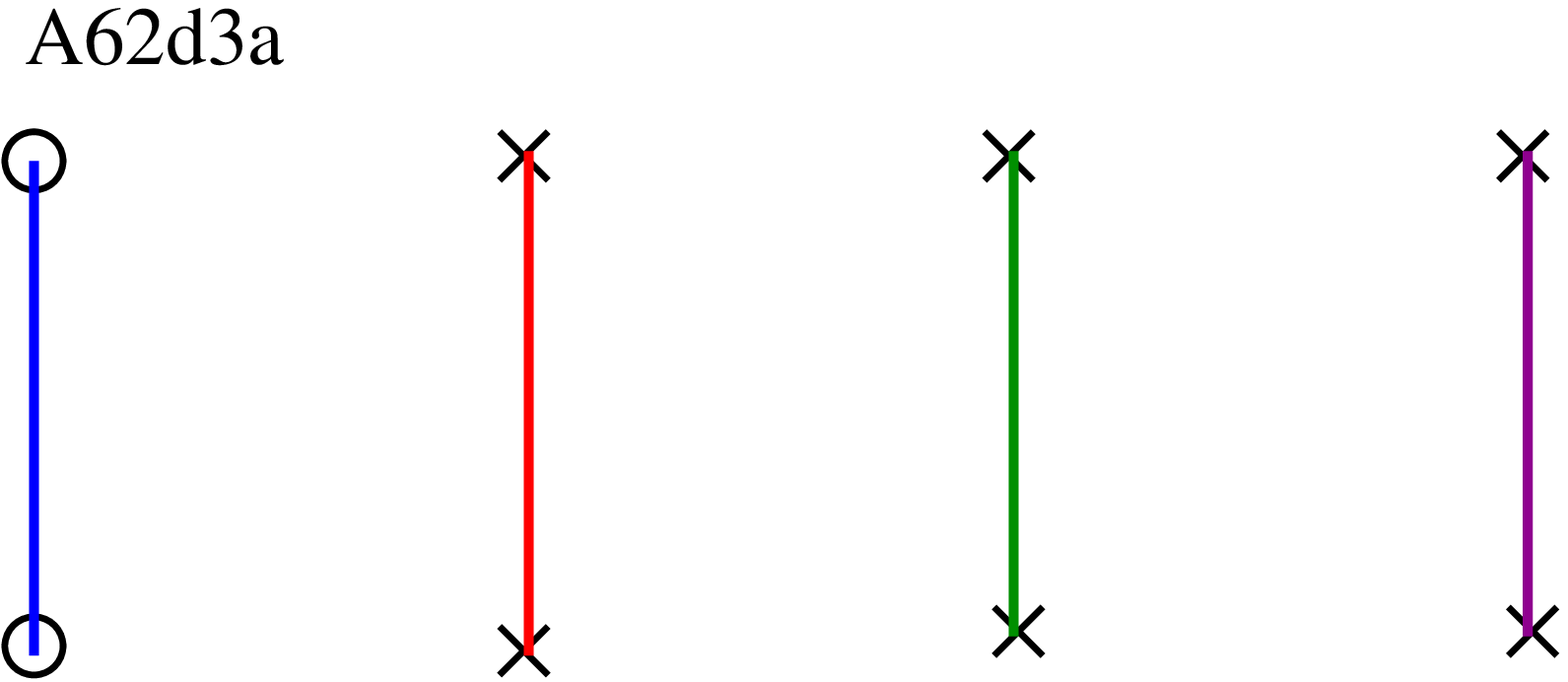}\vspace{0.15in}
\includegraphics[scale=0.135]{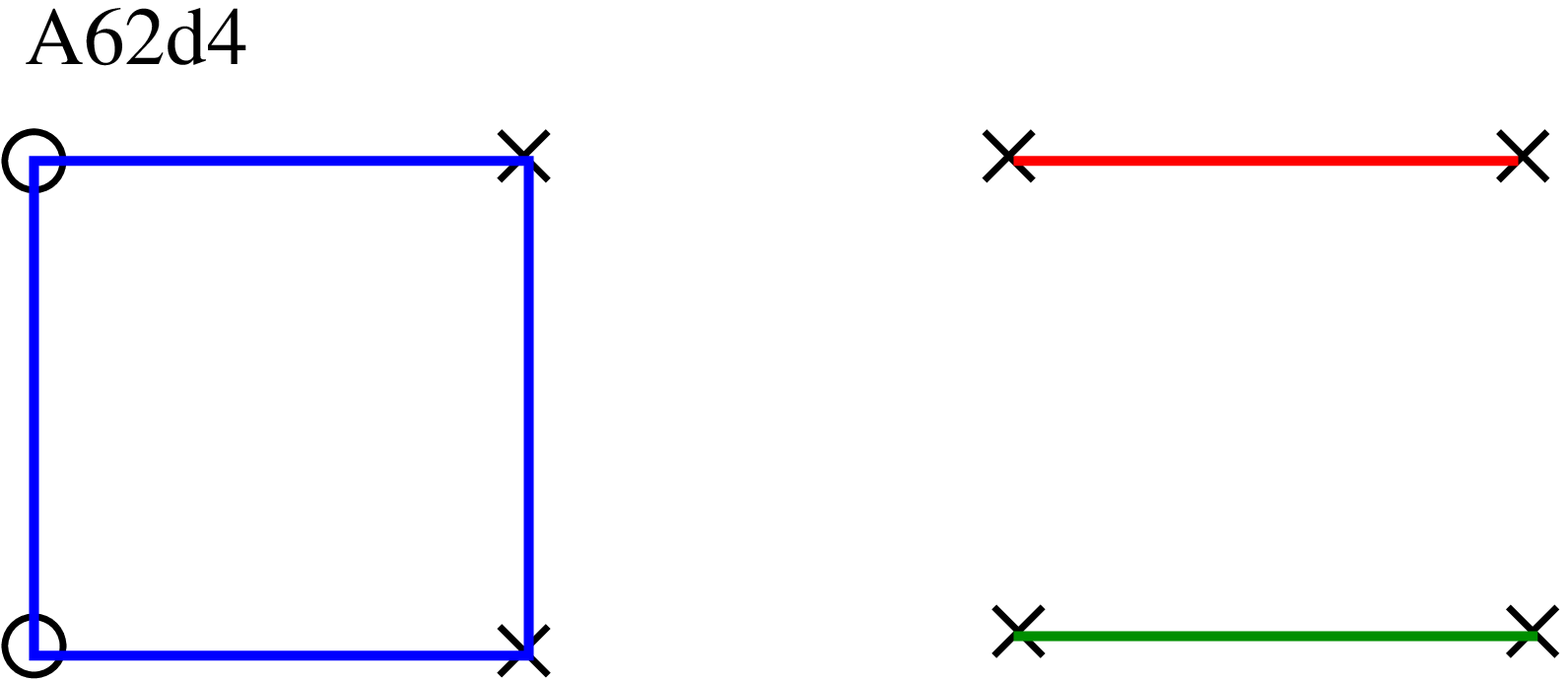}\hspace{0.35in}
\includegraphics[scale=0.135]{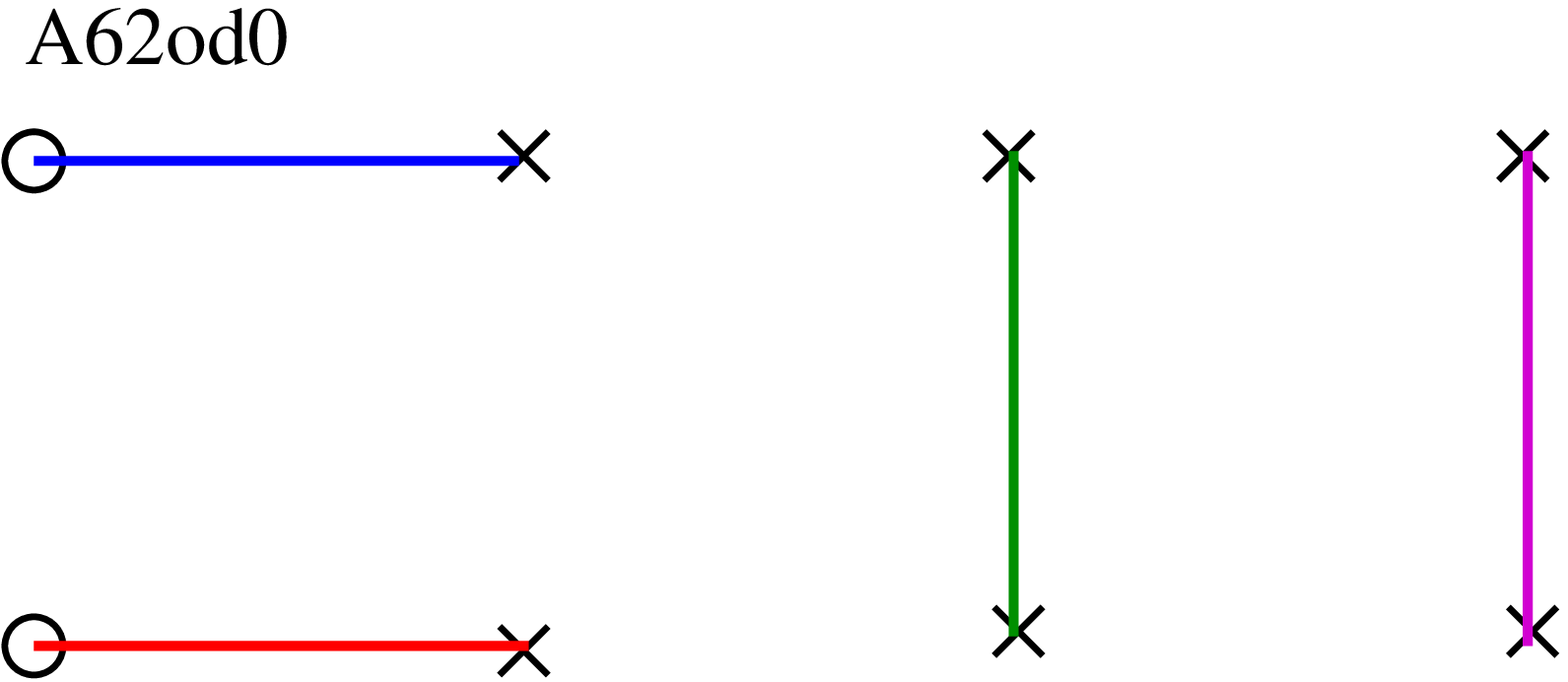}\hspace{0.35in}
\includegraphics[scale=0.135]{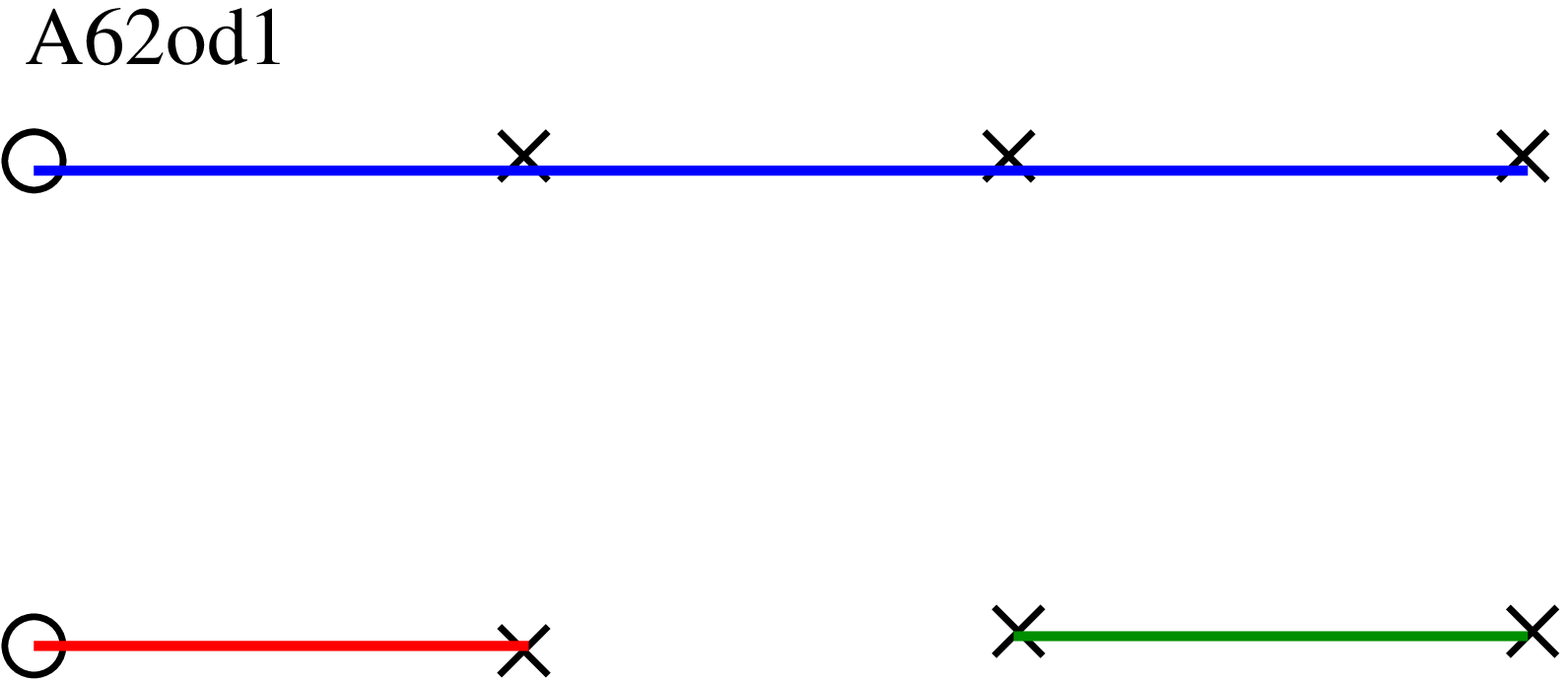}\hspace{0.35in}
\includegraphics[scale=0.135]{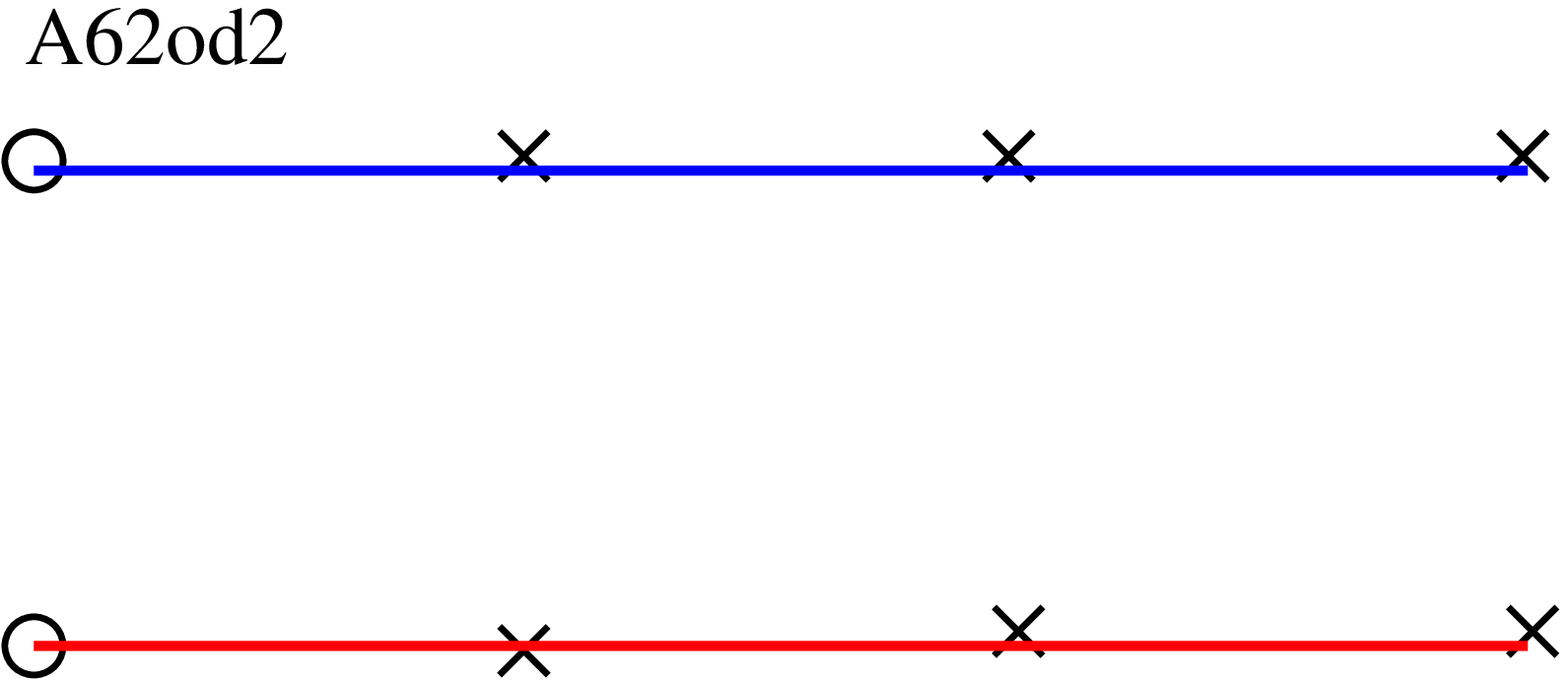}\hspace{0.35in}
\includegraphics[scale=0.135]{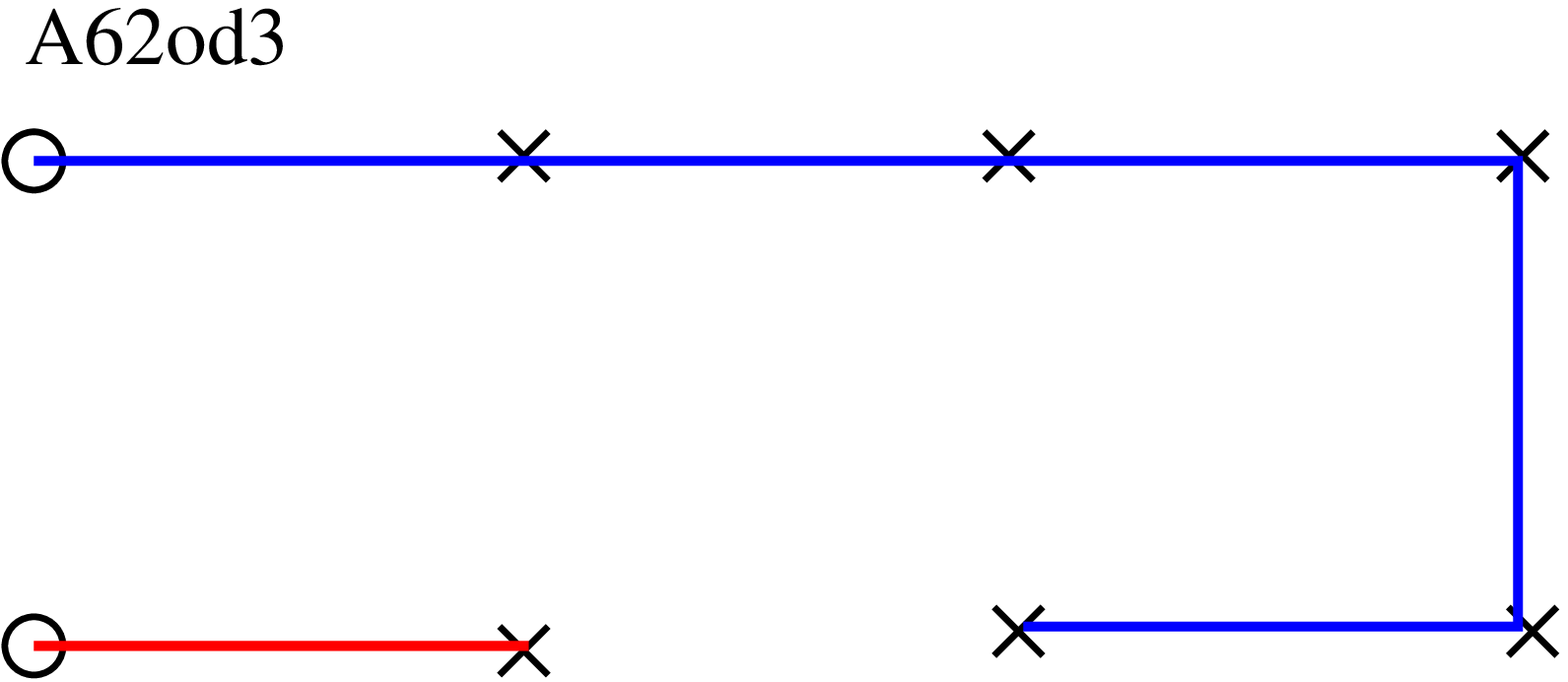}
\end{center}
\caption{\label{fig:A62comb} Graphical depiction of contributing $A^6_2$ index combinations.}
\end{figure}

\begin{figure}
\begin{center}
\includegraphics[scale=0.11]{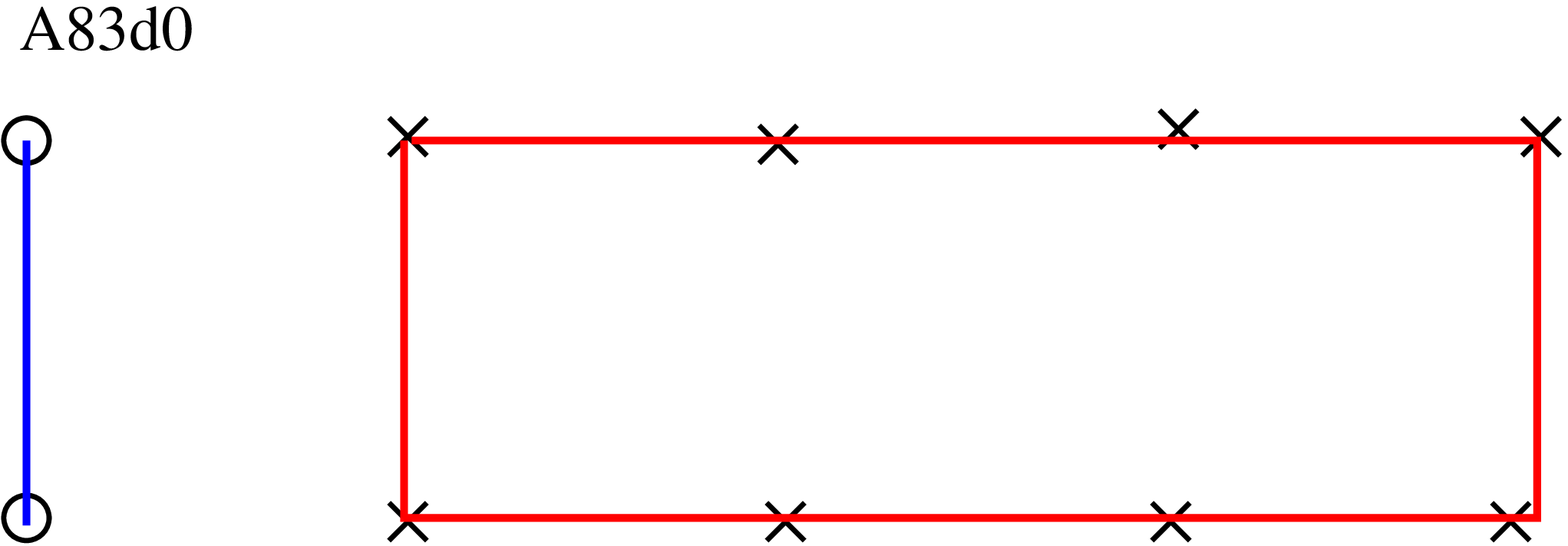}\hspace{0.3in}
\includegraphics[scale=0.11]{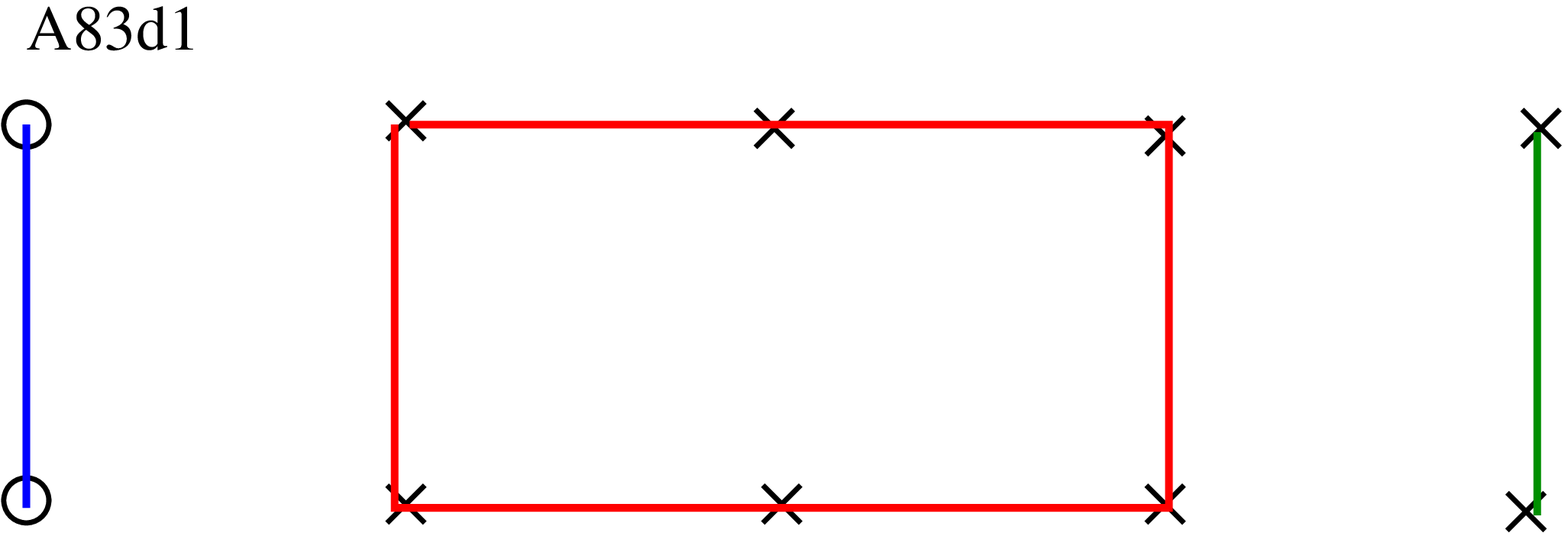}\hspace{0.3in}
\includegraphics[scale=0.11]{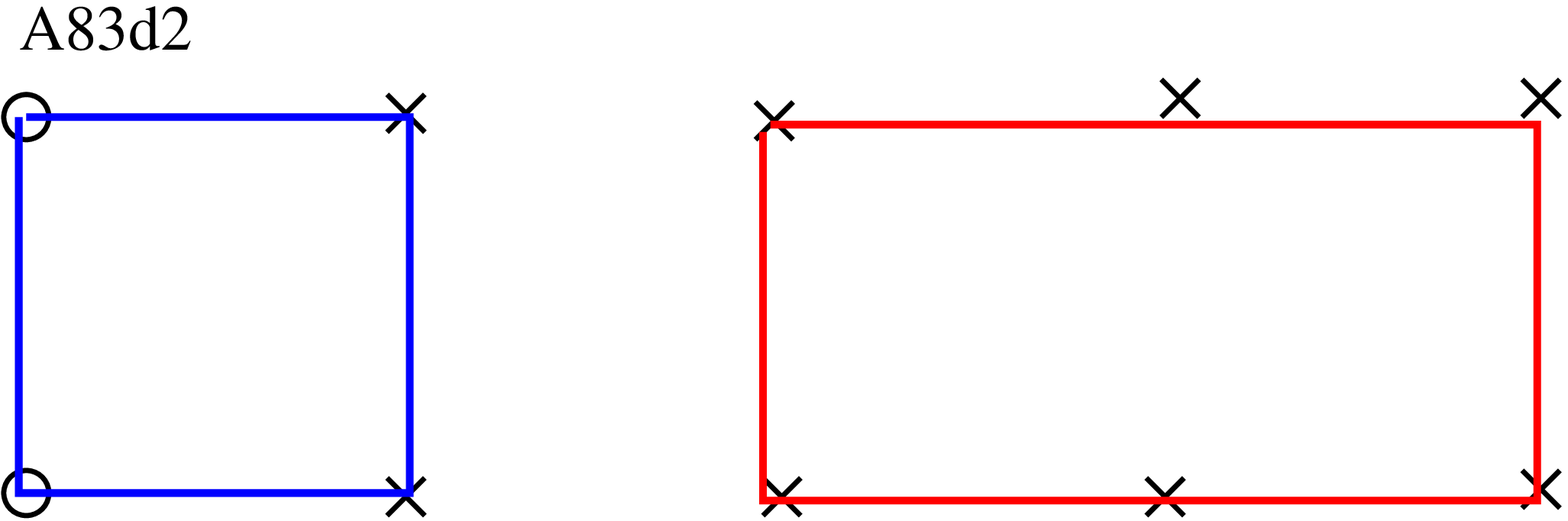}\hspace{0.3in}
\includegraphics[scale=0.11]{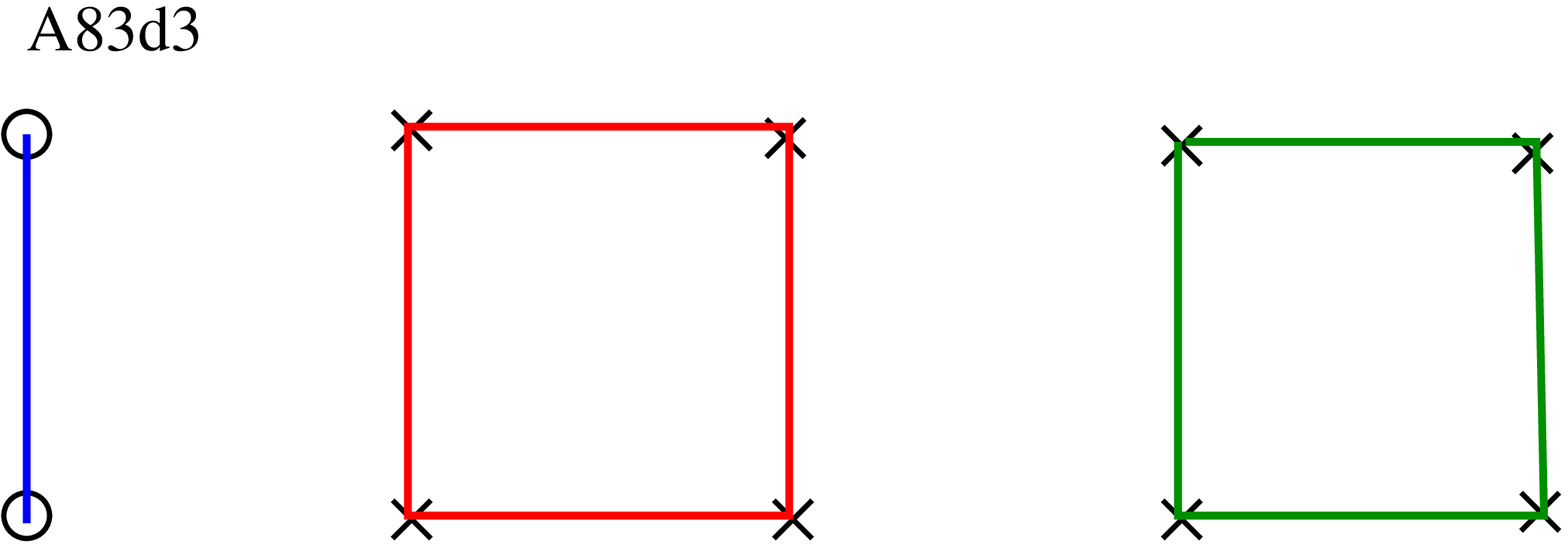}\hspace{0.3in}
\includegraphics[scale=0.11]{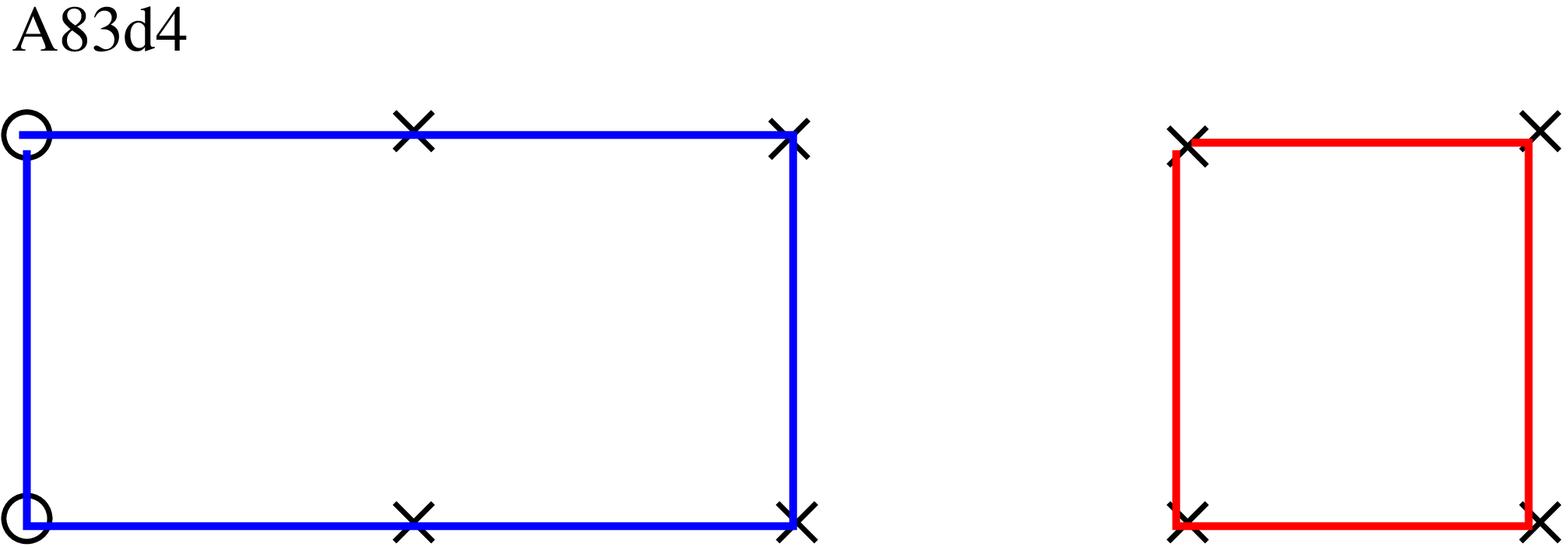}\vspace{0.1in}
\includegraphics[scale=0.11]{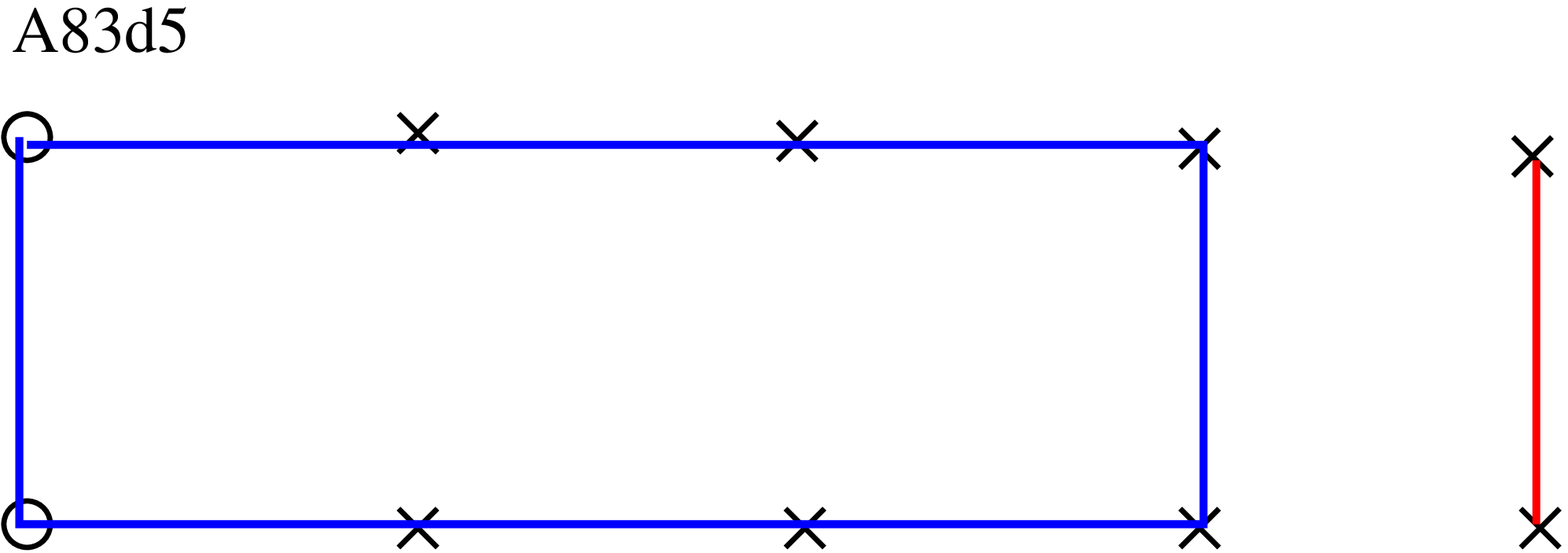}\hspace{0.3in}
\includegraphics[scale=0.11]{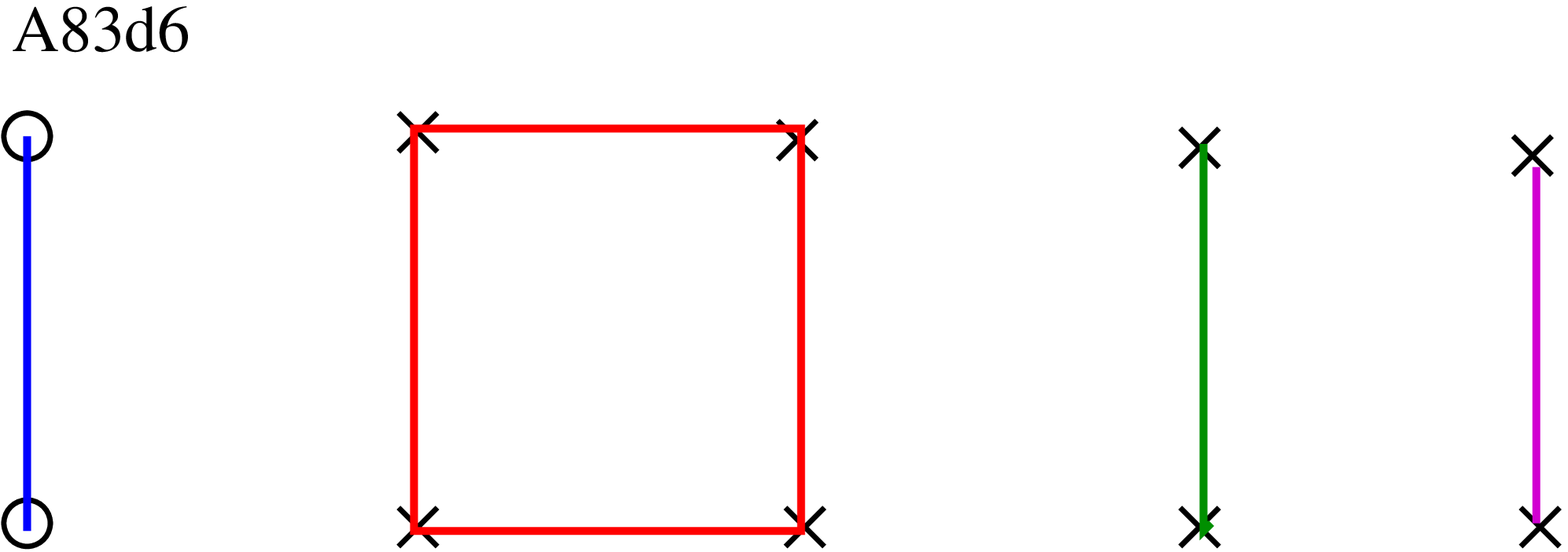}\hspace{0.3in}
\includegraphics[scale=0.11]{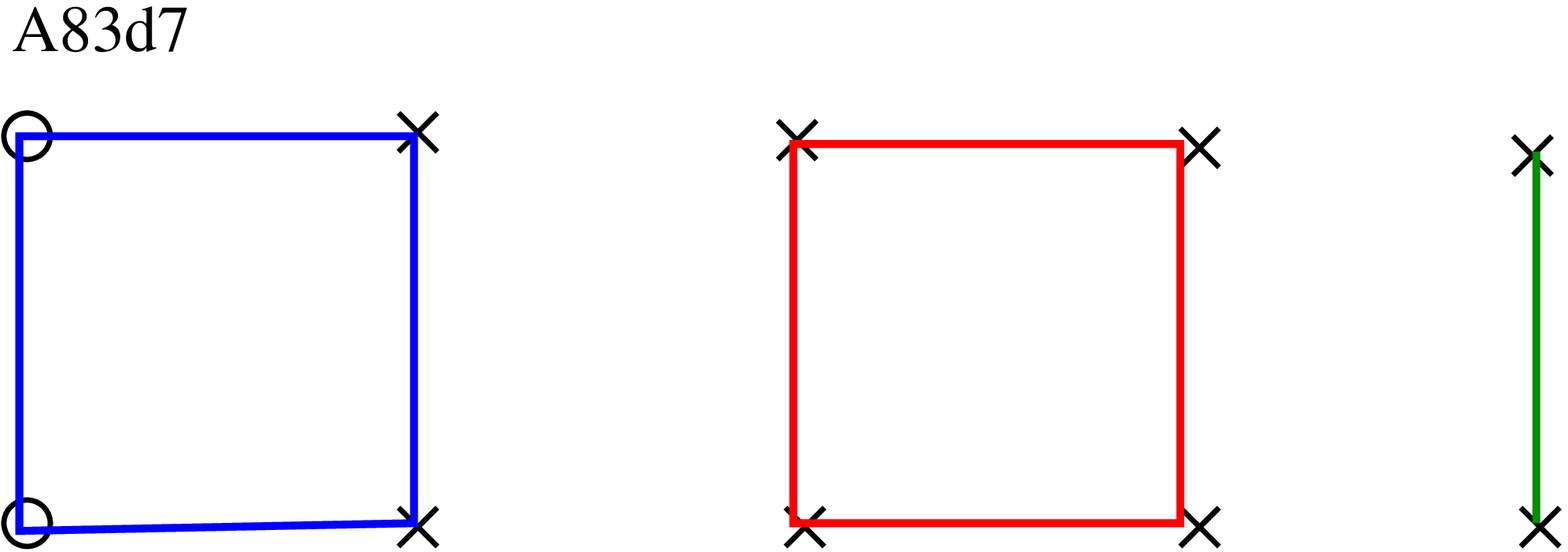}\hspace{0.3in}
\includegraphics[scale=0.11]{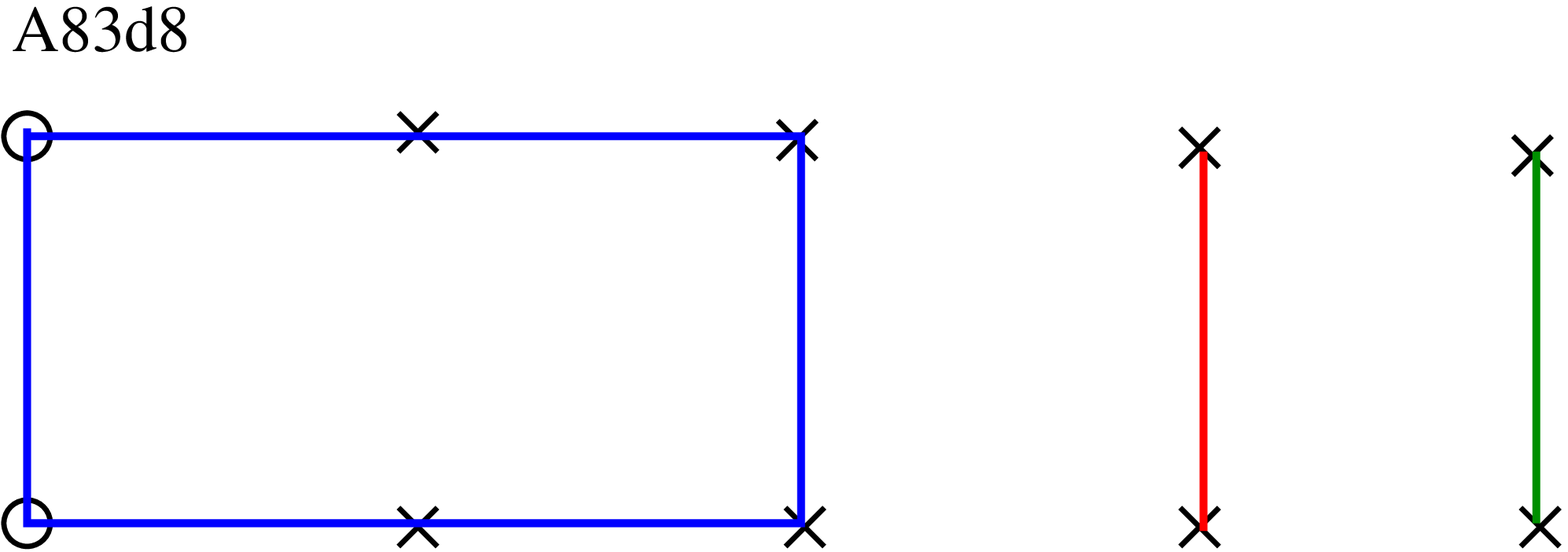}\hspace{0.3in}
\includegraphics[scale=0.11]{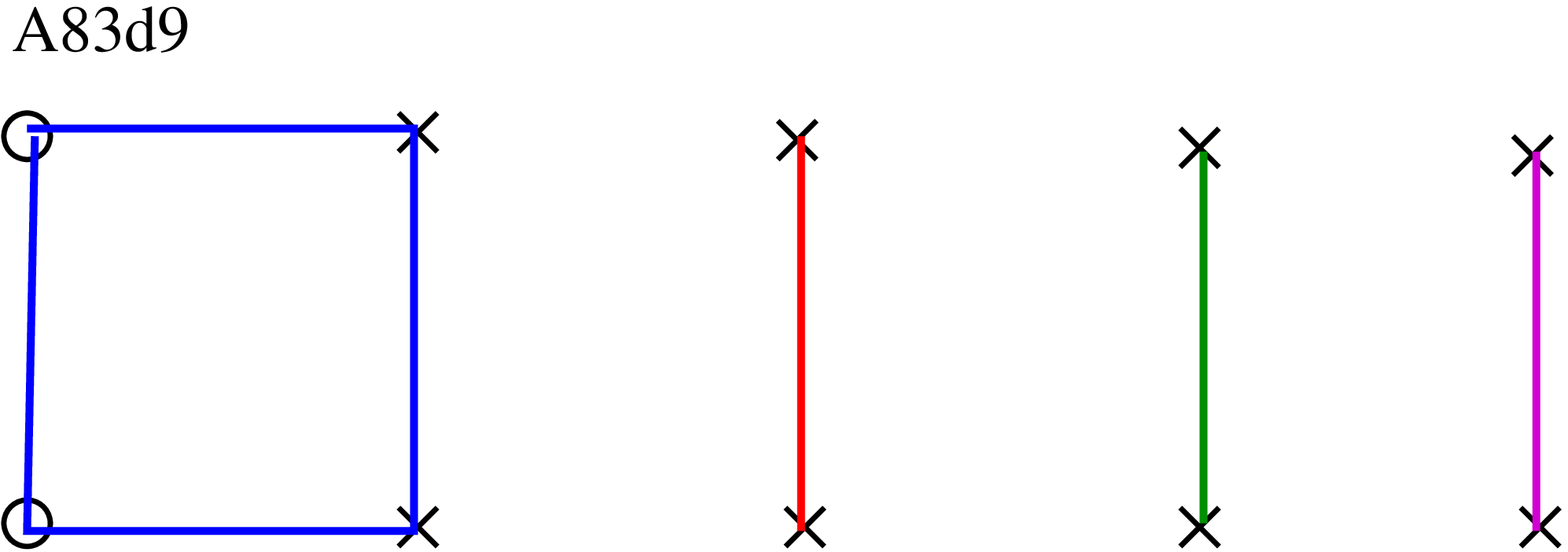}\vspace{0.1in}
\includegraphics[scale=0.11]{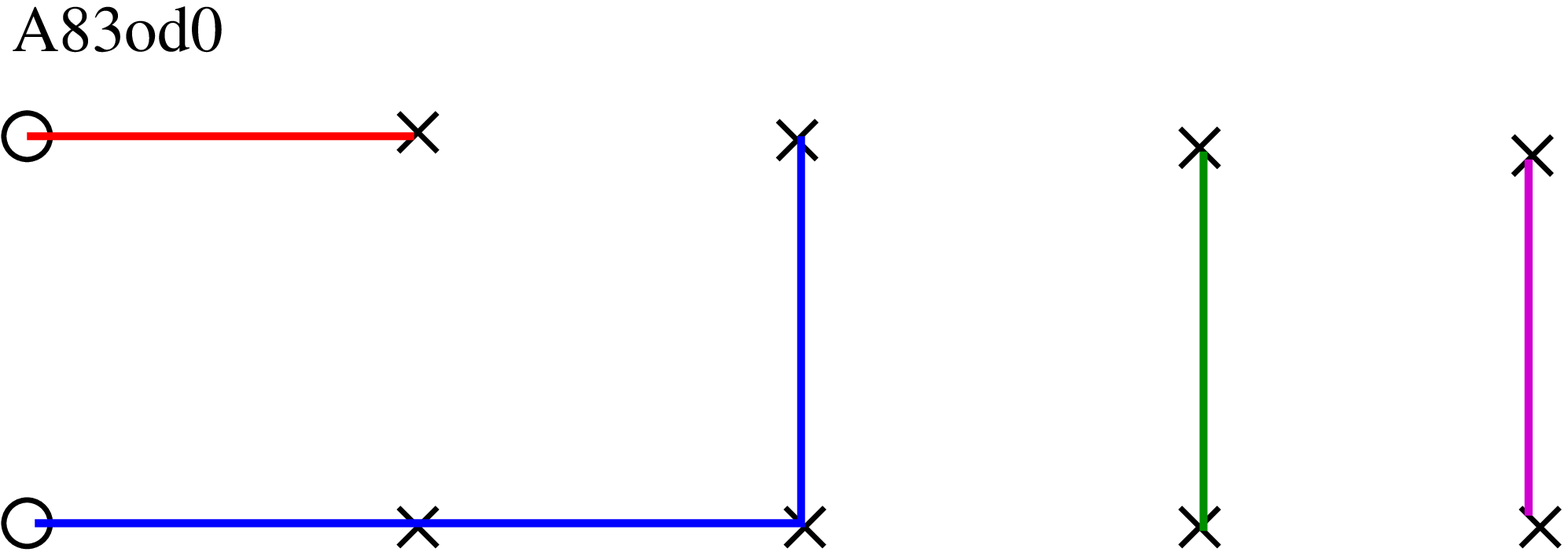}\hspace{0.3in}
\includegraphics[scale=0.11]{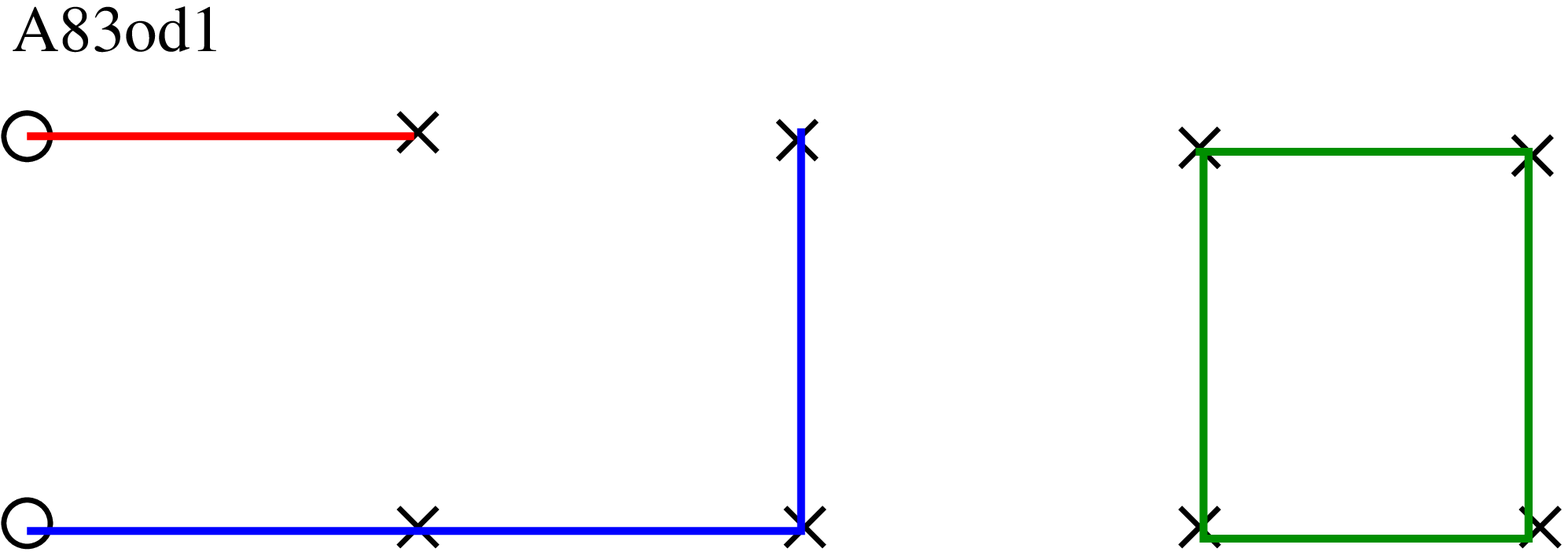}\hspace{0.3in}
\includegraphics[scale=0.11]{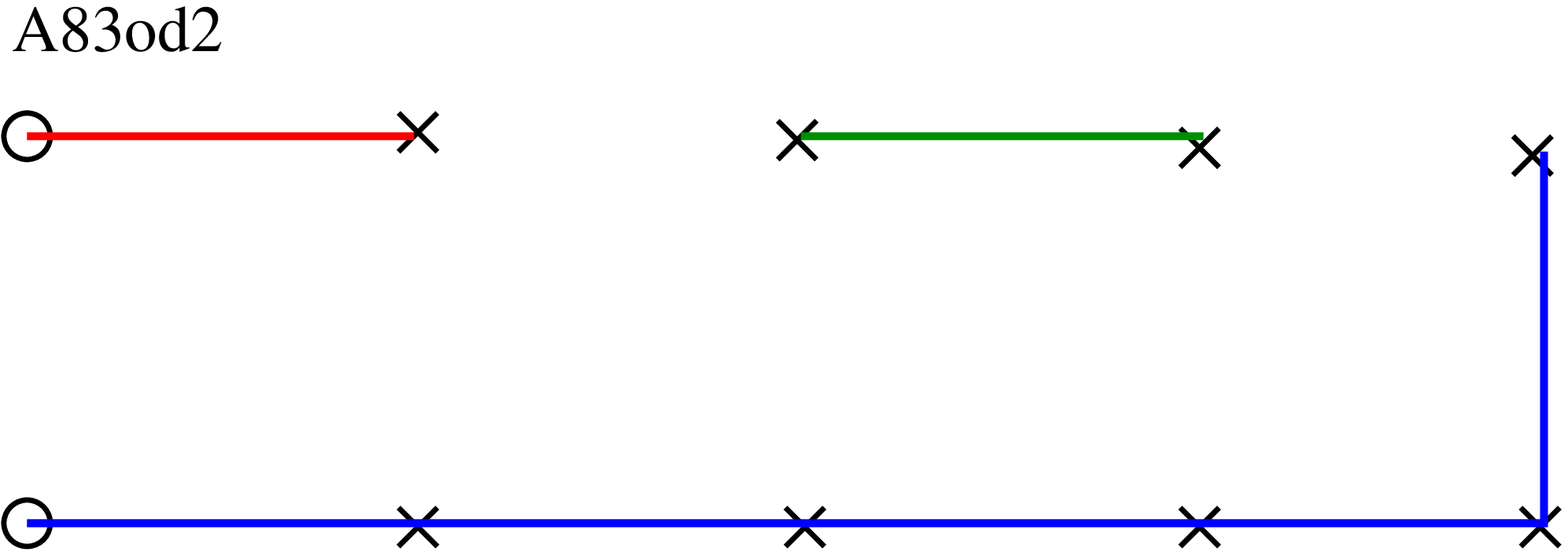}\hspace{0.3in}
\includegraphics[scale=0.11]{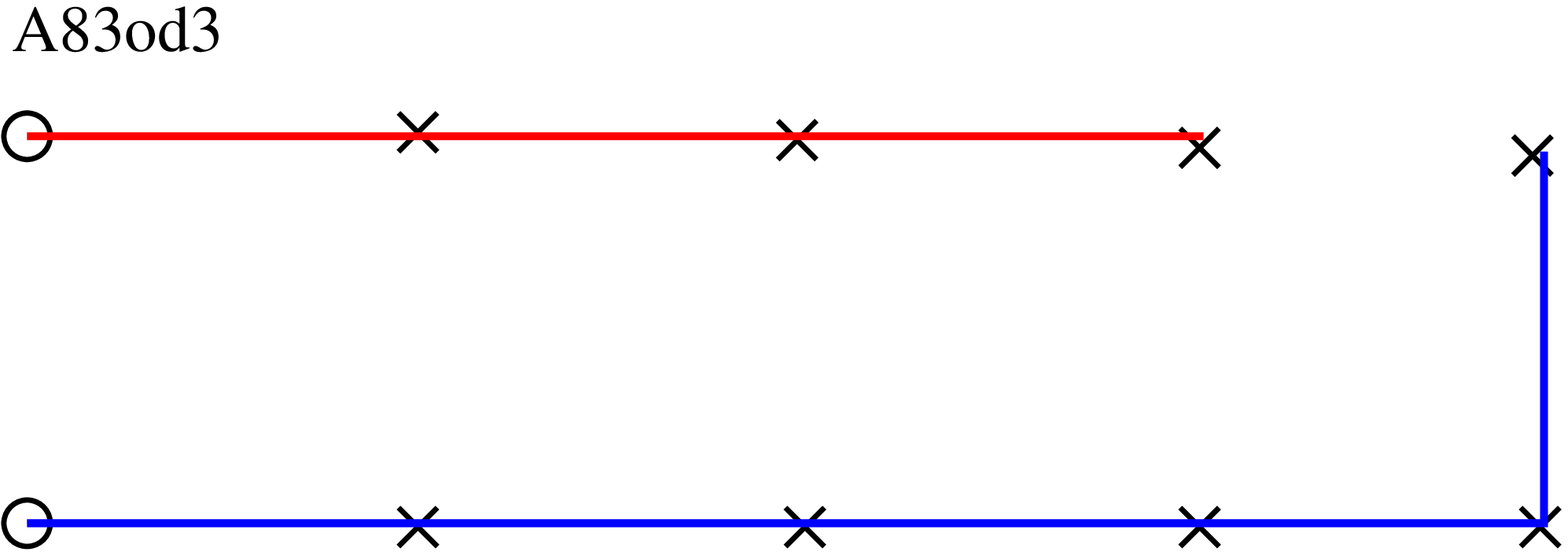}\hspace{0.3in}
\includegraphics[scale=0.11]{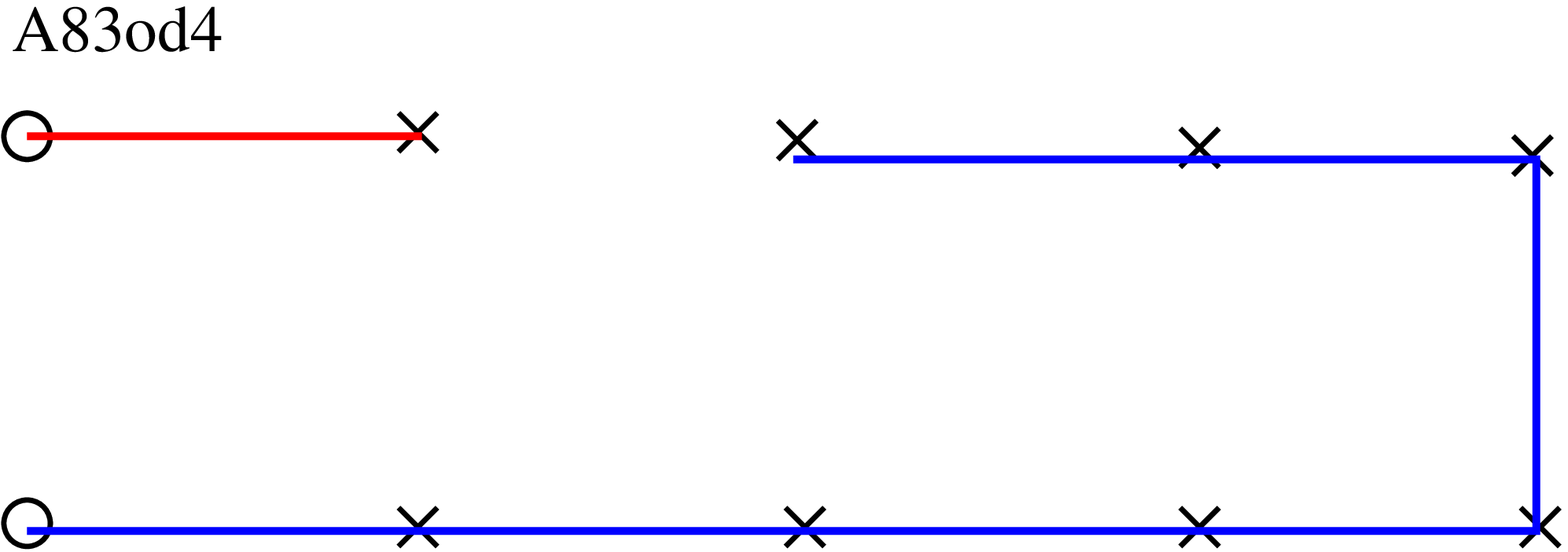}\hspace{0.3in}
\end{center} \vspace{-0.25in}
\caption{\label{fig:A83comb}Graphical depiction of contributing $A^8_3$ index combinations.}
\end{figure}

For $s=0$ we must be slightly more savvy. The factors of $q$ in ${A_{\mu\nu}^{\{\alpha\}}}^n_m$ cause unwanted divergences.
Coefficients ${A_{\mu\nu}^{\{\alpha\}}}^n_m(q)$ have $(2-n)+2m$ powers of momentum.
So for any value of $m$, needed to find the $m^{\rm th}$ derivative of $\Pi(s)$, $n=2+2m$ gives a constant coefficient with no $q$-dependence. Then we can solve
\begin{equation}
\frac{d^m\Pi}{ds^m}\Big|_{s=0}= 
\frac{1}{{A_{\mu\nu}^{\{\alpha\}}}^{(2+2m)}_m}
\frac{\partial^{(2+2m)}\Pi_{\mu\nu}}
{\partial \hat{q}_{\alpha_1}\cdots\partial \hat{q}_{\alpha_{2+2m}} } \Big|_{\hat{q}=0}.
\end{equation}
We concern ourselves with the first three derivatives of $\Pi(s)$. So at $s=0$ the 
relevant coefficients are ${{A_{\mu\nu}}^{2}_0}$, 
${{A_{\mu\nu}}^{4}_2}$, 
${{A_{\mu\nu}}^{6}_1}$, and
${{A_{\mu\nu}}^{8}_3}$.
What remains if to find the cases where the $A^n_m$ are constant for $n=m+2$. For these cases the constants are combinations of Kronecker deltas. 
To make the most of our data we attempt to classify these contributing index combinations.
For $n=2$, $m=0$ we have two cases
\BE
{A_{\mu\nu}^{\{\alpha\}}}^2_0
= (\delta_{{\alpha}_1\mu}\delta_{{\alpha}_2\nu} - 2\delta_{\mu\nu}\delta_{\alpha_1{\alpha}_2\nu})
=
\left\{
\begin{array}{rl}
-2 & \mbox{for $\mu=\nu$, $\alpha_1=\alpha_2$, $\alpha_1\neq\mu$}\\
1  & \mbox{for $\mu=\alpha_1$, $\nu=\alpha_2$, $\mu\neq\nu$}
\end{array}
\right.
\EE
In Tab.~\ref{tab:Amncomb} we summarize the $A^2_0$. 
We label the label diagonal in $\mu$ and $\nu$ as the ``A20d0'' channel. 
There are $N_{\rm comb}=12$ index combinations that contribute. 
If we explore all 
the possible index values for the
off-diagonal $\mu\neq\nu$ case, which we label ``A20od0'', there are $N_{\rm comb}=24$ contributions. However 
 $\alpha_1$ and $\alpha_2$ are interchangeable, so the number of independent second derivatives of $\Pi_{\mu\nu}$ that contribute is smaller. We use a local source at the sink and a conserved vector current (CVC) source at the sink, so $\mu$ and $\nu$ are distinguishable. We therefore have $N_{cl}=12$ combinations for ``A20od0''.  Had we used CVC at both ends we would have only
 $N_{cc}=6$ combinations. We see that in total for our local-CVC setup, we have 24 independent measurements of $\frac{\partial^2 \Pi_{\mu\nu}(0)}{\partial_{{\hat{q}}_{\alpha_1}}\partial_{{\hat{q}}_{\alpha_2}}}$ which contribute to our estimate of $\Pi(s=0)$. 
The contributing index channels for $A^2_0$ are summarized graphically in Fig.~\ref{fig:A20comb}. We classify the contributing channels for $A^4_1$, $A^6_2$, 
and $A^8_3$ in Figs.~\ref{fig:A41comb}, \ref{fig:A62comb} and \ref{fig:A83comb}, respectively. The numbers of contributing independent index configurations 
for each channel of  $A^2_0$, $A^4_1$, $A^6_2$, 
and $A^8_3$ are summarized in Tab.~\ref{tab:Amncomb}

\begin{table}[th]
{\footnotesize
\begin{center}
\begin{tabular}{rrrrr}
$A^2_0$ & label & $N_{\rm comb}$ &$N_{cl}$ & $N_{cc}$\\
\hline
-2 &A20d0 & 12 & 12 & 12\\
1 & A20od0& 24 & 12 & 6\\
\hline
 & total & 36 & 24 & 18\vspace{0.8in}\\

$A^6_2$ & label & $N_{\rm comb}$ &$N_{cl}$ & $N_{cc}$\\
\hline
-360   &\rm{ A62d0}& 12       &  12    & 12  \\
 -72   & \rm{A62d1}& 360      &  24    & 24  \\
 -48   & \rm{A62d2}& 180      &  12    & 12  \\
 -24   & \rm{A62d3}& 180      &  12    & 12  \\
 -24   & \rm{A62d3a}&  360      &  4      &   4   \\
 -16   & \rm{A62d4}& 1080     &  12    & 12   \\
\hline
  +4   & \rm{A62od0}& 2160    &  12    &  6   \\
 +12   & \rm{A62od1}& 3600    &   48   & 24   \\
 +36   & \rm{A62od2}& 240     &   12   &  6   \\
 +60   & \rm{A62od3}& 144     &    24   & 12  \\
\hline  & total     & 8316     &  172  & 124
\end{tabular}
\vspace{0.2in}
\hspace{0.2in}
\begin{tabular}{rrrrr}
$A^4_1$ & label & $N_{\rm comb}$ &$N_{cl}$ & $N_{cc}$\\
\hline
-24   & A41d0 &  12 & 12 & 12\\
 -8   & A41d1 &  72 & 12 & 12\\
 -4   & A41d2 &  72 & 12 & 12\\
\hline
 +2   & A41od0 & 288 & 24 & 12\\
 +6   & A41od1 & 96  & 24 & 12\\
\hline
      & total  & 540    & 84 & 60\vspace{0.2in}\\
$A^6_2$ & label & $N_{\rm comb}$ &$N_{cl}$ & $N_{cc}$\\
\hline
-6720  & \rm{ A83d0}& 12      &    12   & 12    \\
-960  & \rm{ A83d1}& 672      &     24  & 24    \\
-720  & \rm{ A83d2}& 336      &     12  & 12    \\
-576  & \rm{ A83d3}& 840      &     12  & 12    \\
-288   &\rm{ A83d4}& 840      &     12  & 12    \\
 -240  & \rm{A83d5}& 336      &     12  & 12    \\
 -192  & \rm{A83d6}& 5040     &     12  & 12    \\
 -144  & \rm{A83d7}& 10080    &     12  & 12    \\
 -96   & \rm{A83d8}& 5040     &     12  & 12    \\
 -48   & \rm{A83d9}& 10080    &     4   &  4    \\
\hline
  +24   & \rm{A83od0} & 60480 &     24  & 12    \\
 +72    & \rm{A83od1} & 26880 &     48  & 24    \\
 +120   & \rm{A83od2} & 9408  &     48  & 24    \\
 +360   & \rm{A83od3} & 1344  &     24  & 12    \\
 +840   & \rm{A83od4} & 192   &     24  & 12    \\
\hline
        &  total      & 131580 & 292 & 208
\end{tabular}
\end{center}
}\vspace{-0.4in}
\caption{\label{tab:Amncomb}Combinations contributing to non-zero $A^2_0$, $A^2_0$,
 $A^2_0$ and  $A^2_0$. }
\end{table}


\subsection{Smooth curve generation}
For  $s$ between two lattice momenta $s_i < s < s_{i+1}$, 
we make  ``lower'' and ``upper'' estimates,
\begin{equation}
\Pi^{\rm low}(s) =
\sum_{n}\left(s - s_{i}\right)^n \frac{1}{n!}\frac{d^n \Pi}{ds^n}\Big|_{s_{i}}
{\;\;\;\;\;\;\;\;\;}
{\rm and}
{\;\;\;\;\;\;\;\;\;}
\Pi^{\rm up}(s) =
\sum_{n}\left(s - s_{i+1}\right)^n \frac{1}{n!}\frac{d^n \Pi}{ds^n}\Big|_{s_{i+1}}
\end{equation}
We combine these in a weighted average to get a smooth function $\Pi^{\rm sm}$ for
the integrand of (\ref{integral}).
\begin{equation}
\Pi^{\rm sm}_p(s) = \frac{\Pi^{\rm low}(s)w^{\rm low}(s) + \Pi^{\rm up}(s)w^{\rm up}(s)}
{w^{\rm low}(s)+w^{\rm up}(s)}
\end{equation}
with
\begin{equation}
w^{\rm low}(s) = \frac{1}{\left|\left(s - s_{i}\right)\sigma\left(\frac{d\Pi}{ds}\Big|_{s_{i}}\right)\right|^{p}}
{\;\;\;\;}
{\rm and}
{\;\;\;\;}
w^{\rm up}(s) = \frac{1}{\left|\left(s - s_{i+1}\right)\sigma\left(\frac{d\Pi}{ds}\Big|_{s_{i+1}}\right)\right|^{p}}.
\end{equation}
$\sigma\left(\frac{d\Pi}{ds}\right)$ is a proxy for the uncertainty in $\Pi^{\rm low/up}$ and $p$ is an adjustable parameter.

\section{Numerical tests}

\begin{figure}
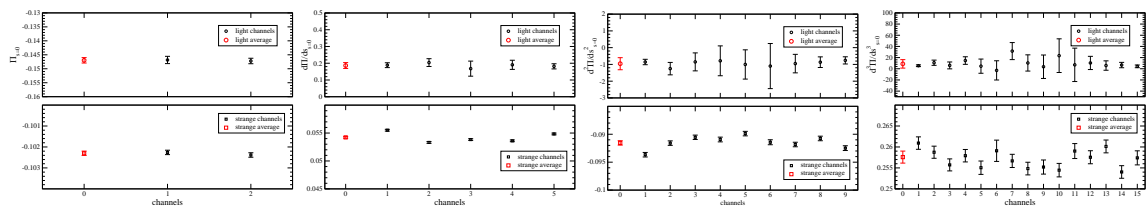

\begin{center}
\includegraphics[scale=0.155]{GRAPHICS/d0Pi_channels_b3.5_m0.05294_m0.006.eps}
\includegraphics[scale=0.155]{GRAPHICS/d1Pi_channels_b3.5_m0.05294_m0.006.eps}
\includegraphics[scale=0.155]{GRAPHICS/d2Pi_channels_b3.5_m0.05294_m0.006.eps}
\includegraphics[scale=0.155]{GRAPHICS/d3Pi_channels_b3.5_m0.05294_m0.006.eps}
\end{center}\vspace{-0.2in}
\caption{\label{fig:channels}Test of consistency of estimates of $\frac{d^m\Pi}{ds^m}$
from different channels.
}
\end{figure}

\begin{figure}
\begin{center}
\includegraphics[scale=0.2]{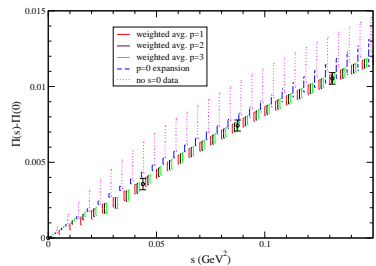}
\end{center}
\vspace{-0.2in}
\caption{\label{fig:smoothcurves}Smooth $\Pi(s)$ curves generated with different values of $p$. }
\end{figure}

\begin{figure}
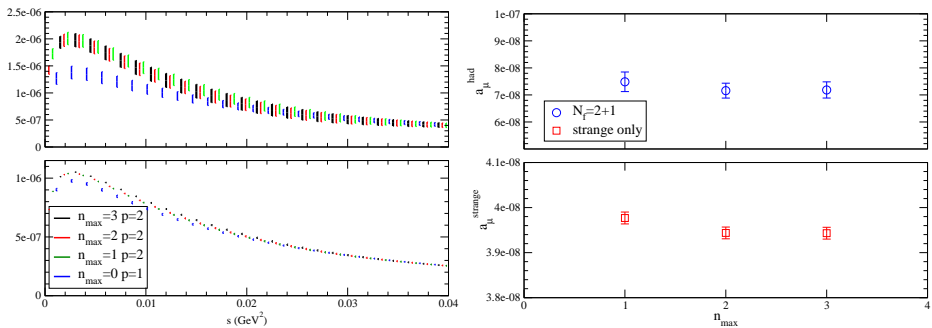

\begin{center}
\includegraphics[scale=0.25]{GRAPHICS/integrand_hex2_b3.5_m0.05294_m0.006_64_64_nmax_combi.eps}
\includegraphics[scale=0.25]{GRAPHICS/integral_nmax_scan_p2_hex2_b3.5_m0.05294_m0.006_64_64_combi.eps}
\end{center}
\vspace{-0.2in}
\caption{\label{fig:ntest}The dependence on the maximum expansion order $n$ of the integrand (l) and $a_{\rm HVP}$  (r).}
\end{figure}

We have tested this method on several of the $N_f = 2+1$ flavor 2-HEX ensembles
from BMW-c~\cite{Durr:2008zz}. 
For this work we concentrate on the ensemble listed in  Tab.~\ref{tab:configs}, which has the advantage of having $1060$ configurations and $L_s=L_t$. The strange quark mass is mis-tuned on this ensemble, so the data from
the additional ensembles is needed to correct for it.
We show in Fig.~\ref{fig:channels} that the different 
channels for each $A^n_m$ yield consistent estimates of $\frac{d^m\Pi}{ds^m}$. 
In Fig.~\ref{fig:smoothcurves} we test different methods of computing a smooth function 
of $\Pi$, including different values of $p$. We note as a curiosity, the large error that would be induced by neglecting the $s=0$ point, and how well one might do using {\em only} the $s=0$ point. Fig.~\ref{fig:ntest} demonstrates that $n=3$ is a sufficient expansion order for determining a smooth function $\Pi$.

\begin{table}[htb]
\begin{center}
{\footnotesize
\begin{tabular*}{\textwidth}{c @{\extracolsep{\fill}}cccc}
$am_{ud}^{\rm bare}$ & $am_s^{\rm bare}$ & volume & \#\,cfgs\ & $M_\pi$ (GeV) \\
\hline
\multicolumn{5}{c}{$\beta = 3.5$, $a^{-1} = 2.131$ GeV}\\
  -0.05294 & -0.0060 & $64^3\times 64$ & 1060 & 0.130(2)\\
\end{tabular*}
}
\end{center}\vspace{-0.3in}
\caption{\label{tab:configs}Configuration parameters.}
\end{table}

\vspace{-0.1in}

\section{Conclusions}

The method described above
uses many estimates of the  spatial and  temporal moments
to make a precise determine of $\Pi(s)$ and its derivatives at both
finite and zero momentum. Additional systematic errors need to be studied
such as finite volume effects~\cite{Malak:2015sla}.

Including spatial as well as temporal moments greatly increases the
number of estimates of $\Pi$ and its derivatives at $s=0$ one can
obtain from each source on each configuration. 
The $s=0$ point is the most important in the determination of $a_{\rm HVP}$, because
it is so much closer to the peak of the integrand in equation~\ref{integral}, than the 
first finite $s$ lattice momentum available for current lattice volumes. The most 
important lattice measurement one can make for determining $a^{\rm HVP, LO}_{\mu}$ is 
$\frac{d\Pi}{ds}\Big|_{s=0}$, because $\Pi(0)$ is subtracted off. Our method 
 produces 172 estimates of $\frac{d\Pi}{ds}\Big|_{s=0}$ for each source.

The authors thank
the Gauss Centre for Supercomputing (GCS) 
for providing computing time through the John von Neumann Institute
for 
Computing (NIC) on the GCS share of the supercomputer JUQUEEN at
J\"ulich 
Supercomputing Centre (JSC) and time granted on 
JUROPA at JSC.
This work used the DiRAC Blue Gene Q Shared Petaflop system at the
University of Edinburgh,
operated by the Edinburgh Parallel Computing Centre on behalf of the
STFC DiRAC HPC Facility 
(www.dirac.ac.uk). 

\bibliographystyle{h-physrev5}
\bibliography{latt2015}

\end{document}